\documentclass[onecolumn,draftcls,12pt,journal]{IEEEtran}
\usepackage{graphicx}
\usepackage{amsfonts}
\usepackage{mathrsfs}
\usepackage{amssymb,amsmath}
\usepackage{bm}
\usepackage{algorithm}
\usepackage{algorithmic}
\usepackage{theorem}
\usepackage{array,color}
\usepackage{cite}
\usepackage{stfloats}
\usepackage{subfigure}
\usepackage{array}
\usepackage{multirow}
\usepackage{bbding}

\newcommand{\rd}{\textcolor{black}}
\newcommand{\edit}[1]{\textcolor{black}{#1}} 
\newcommand{\ist}{\hspace*{.3mm}}
\newcommand{\rmv}{\hspace*{-.3mm}}
\newcommand{\response}{\textcolor{black}}

\begin{document}
\title{Cooperative Joint Localization and Clock Synchronization Based on Gaussian Message Passing in Asynchronous Wireless Networks}
\author{Weijie Yuan, Nan Wu, Bernhard Etzlinger, Hua Wang and Jingming Kuang
\thanks{This work is supported by the ``National Science Foundation of China
(NSFC)'' (Grant No. 61201181) and ``A Foundation for the Author of National Excellent Doctoral Dissertation of P. R. China (FANEDD)'' (Grant No. 201445).

W. Yuan, N. Wu, H. Wang and J. Kuang are with the School of Information and Electronics, Beijing Institute of Technology, Beijing, China (Email: \{wjyuan, wunan, wanghua, jmkuang\}@bit.edu.cn).

B. Etzlinger is with the the Institute of Communications Engineering and RF-Systems, Johannes Kepler University, Linz, Austria (Email: b.etzlinger@nthfs.jku.at).

Corresponding author: Nan Wu}
}

\maketitle
\begin{abstract}
Localization and synchronization are very important in many wireless applications such as monitoring and vehicle tracking. Utilizing the same time of arrival (TOA) measurements for simultaneous localization and synchronization is challenging. In this paper, we present a factor graph (FG) representation of the joint localization and time synchronization problem based on TOA measurements, \rd{in which the non-line-of-sight measurements are also taken into consideration}. On this FG, belief propagation (BP) message passing and variational message passing (VMP) are applied to derive two fully distributed cooperative algorithms with low computational requirements. Due to the nonlinearity in the observation function, it is intractable to compute the messages in closed form and most existing solutions rely on Monte Carlo methods, e.g., particle filtering. We linearize a specific nonlinear term in the expressions of messages, which enables us to use a Gaussian representation for all messages. Accordingly, only the mean and variance have to be updated and transmitted between neighboring nodes, which significantly reduces the communication overhead and computational complexity. A message passing schedule scheme is proposed to trade off between estimation performance and communication overhead. Simulation results show that the proposed algorithms perform very close to particle-based methods with much lower complexity especially in densely connected networks.
\end{abstract}
\section{Introduction}
Wireless networks play a main role in the modern societies. For many applications of wireless networks such as public service, emergence rescue and intelligent vehicular system, position information is a crucial requirement for the network to function as intended\cite{gezici2005localization}. Generally, the Global Positioning System (GPS) can provide accurate location in our daily life. However, equipping all wireless nodes (e.g., sensors, vehicles, people) with GPS receivers may be cost and energy prohibitive. Furthermore, the poor signal penetration capabilities of the widely used GPS lead to inadequate location information\cite{misra2006global}. A cooperative localization algorithm\cite{patwari2005locating} that enables ranging and position information exchange between neighboring nodes can overcome this problem. During the last ten years, there are many research papers focused on cooperative localization algorithms in wireless sensor networks, vehicular networks and acoustic sensor networks{\cite{tseng2009hybrid ,wyme
 ersch2009cooperative,nguyen2015least,6476037,vakulya2011fast}}.

Those cooperative localization methods are based on
\edit{a set of nodes with known locations and on a set of range measurements among neighboring nodes.}
\response{The range measurements can be obtained using time of arrival (TOA) \cite{chan2006time}, }
time difference of arrival (TDOA) \cite{gustafsson2003positioning},
\edit{round-trip time of arrival (RTT)} \cite{vossiek2003wireless}
or received signal strength (RSS) \cite{del2003link} \edit{measurements}. 
\edit{RSS measurements have the drawback of being sensitive to changes in the environment, whereas the time-based methods alleviate this problem.
Moreover, since TDOA and RTT mechanisms may not be supported by many communication protocols, we will focus on TOA-based techniques.}
However,
\edit{utilizing TOA measurements to obtain accurate range estimates is difficult in the presence of time offsets among the nodes}.
Hence, clock synchronization is a vital requirement in
TOA based
localization
\edit{methods}.
Various synchronization algorithms have been proposed in the literature. In \cite{ganeriwal2003timing} and \cite{maroti2004flooding}, two synchronization protocols, namely, Timing-sync Protocol for Sensor Networks (TPSN) and Flooding Time Synchronization Protocol (FTSP), are proposed. In \cite{schenato2007distributed}, consensus algorithms are used to synchronize all the nodes to the same virtual clock. A factor graph (FG) based distributed network synchronization algorithm using belief propagation is proposed in \cite{leng2011distributed}. An extension to mean-field message passing for cooperative synchronization algorithm is obtained in \cite{etzlinger2013cooperative},
\edit{which}
has the advantage of broadcasting information to neighboring nodes.

The literature above treat time synchronization independent of the localization task. However, the two problems are closely related and it is possible to explore a joint estimation method. Furthermore, in a harsh or mobile environment, the clock of nodes varies and re-synchronization between nodes frequently increases the energy consumption. Recently, based on the closed relationship between the problems of localization and synchronization, several research works have studied simultaneous estimation of positions and clock information of nodes. In \cite{denis2006joint}, the two problems are solved together by performing time synchronization first and then localization, which is not a strict simultaneous approach. The joint time synchronization and localization problem with accurate and inaccurate anchors have been solved in \cite{zheng2010joint} using least squares (LS) and generalized total least squares (GTLS) methods, which is a hierarchical protocol that poses strong topological constraints on the network. A closed-form solution of joint estimation using weighted least squares (WLS) is proposed in \cite{zhu2010joint}.\,In \cite{huang2013efficient}, the authors extend Bancroft's algorithm\cite{bancroft1985algebraic} to overcome the problem of solution ambiguity using LS criterion. An expectation-maximization (EM) based algorithm which recursively estimates the clock parameter and position of unknown node is presented in \cite{ahmad2013joint}.

However, in the above methods, only one node to be synchronized and located is considered, which is different to the situation in cooperative localization where nodes help each other to achieve self-localization and network synchronization. In \cite{meyer2013distributed} and \cite{etzlinger2013cooperative1}, a particle-based belief propagation
\edit{(BP) algorithm}
and hybrid message passing algorithm,
\edit{respectively, have been proposed }
for cooperative simultaneous localization and
\edit{synchronization.}
\edit{Although the algorithms are fully distributed and enable synchronization and localization of multiple nodes, they rely on particle filtering to deal with nonlinear expressions in the message computation.}

\rd{Moreover, the non-line-of-sight (NLOS) propagation in indoor environments can delay the TOA even when the whole network is synchronous, which leads to a positively biased range measurement \cite{Guvenc}. The NLOS in the localization problem has been investigated in several papers. In \cite{gezici2004uwb}, the NLOS problem in UWB signaling is considered. Sum-product algorithm and expectation propagation based on particle filtering for cooperative localization in mixed LOS/NLOS environment is studied in \cite{van2012comparison}. A machine learning approach is proposed in \cite{wymeersch2012machine} for NLOS propagation identification. An analysis of NLOS conditions in wireless localization is performed in \cite{liu2013analysis}. However, to the best knowledge of the authors, joint localization and clock synchronization considering NLOS propagation has not been studied.}

In this paper, we consider \edit{a} two-dimensional localization problem based on TOA measurements in \edit{an} asynchronous wireless network {with clock offset} \edit{among the nodes}.\footnote{\edit{The spatial extension to the three-dimensional case and the temporal extension to asynchronity in both clock phase and frequency offsets are straightforward.}} \edit{We present a FG representation, on which BP message passing \cite{pearl1986fusion} and variational message passing (VMP) \cite{winn2005variational} are applied to derive two joint synchronization and localization algorithms in both LOS and NLOS environments.} Taylor expansions have been used to linearize the nonlinear term in the observation function. With the approximations, all messages on FG can be represented in closed Gaussian form. Thus, the means and variances of beliefs of nodes' estimates can be easily obtained by multiplication and addition operations. The complexity and communication requirements of our proposed algorithm is much lower than that of the particle based algorithms.
Moreover a message passing schedule scheme \response{in which nodes perform more than one internal iterations to update the outgoing messages before they transmit them to neighbors} is presented to further reduce the number of message exchanges between nodes and therefore reduce the communication overhead.

The rest of the paper is organized as follows. The system model is given in Section II. In Section III, the two message passing algorithms for joint localization and synchronization are proposed. The message passing schedules are presented. Simulation results and discussions are given in Section IV. Conclusions are drawn in Section V.

\section{System Model}
We consider a dynamic network comprising
\edit{a set $\mathcal{M} = \{1, ..., M\}$ of}
agent nodes to be located and synchronized, and
\edit{a set $\mathcal{A} = \{1, ..., A\}$ of}
anchor nodes with fixed known positions and timings, where the location and time offset of node $i\in{\cal M}\cup{\cal A}$ at time slot $n$ is denoted by $\bm{x}_{i,n}=[x_{i,n},y_{i,n}]^T$ and $\theta_{i,n}$, respectively. {Herein the agent nodes can be sensors in wireless sensor networks or vehicle devices in vehicular networks.} In the considered system, it is assumed that all anchor nodes are synchronized with the same reference time, that is to say the time offset $\theta_{i}=0,~\forall i\in\cal A$.

The local clock time of node $i\in{\cal M \edit{\cup{\cal A}}}$ is
\begin{align}\label{localclock}
c_{i}(t_i)= \edit{t}+\theta_i \,,
\end{align}
\edit{where $t$}
is the accurate reference time.
\edit{If a}
node $i\in{\cal M}\cup{\cal A}$ is able to exchange information with
\edit{a node $j\in \{{\cal M} \cup {\cal A}\}  \setminus \{ i \}$ at time $n$, the pair $(i,j)$ is collected in the communication set ${\Xi}$.}
\edit{We further collect all $j$ for which $(i,j) \in {\Xi}$ in the neighbor set $\mathcal{S}_{i,n}$ of node $i$. Thus, node $i$ has $N(i) = |\mathcal{S}_{i,n}|$ neighbors at time $n$.}
The sets ${\cal M}_{i,n}={\cal M}\cap {\mathcal S}_{i,n}$ and ${\cal A}_{i,n}={\cal A}\cap {\mathcal S}_{i,n}$
\edit{denote}
neighboring agent nodes and neighboring anchor nodes of node $i$ at time $n$,
\edit{respectively}.


\begin{figure}
\centering
\includegraphics[width=.7\textwidth]{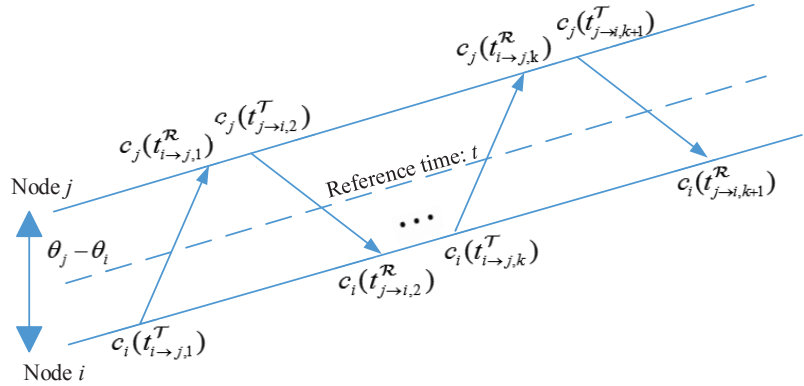}
\caption{\quad TOA time stamp model. The dashed line stands for the reference time and the solid lines are the recorded time stamps at node $i$ and node $j$ with clock offsets.
}\label{Fig1}
\centering
\end{figure}

As shown in Fig. \ref{Fig1}, at time $t^{\mathcal {T}}_{i\to j,n}$, node $i$ transmits its current timing information to node $j$. After a delay $\Delta_{ij,n}$, node $j$ receives the timing information from node $i$ at time $t^{\mathcal R}_{i\to j,n}${, where the subscript $n$ is the time stamp index} \response{and the superscript $\mathcal T$ and $\mathcal R$ are short for ``transmitter side'' and ``receiver side''}. The delay $\Delta_{ij,n}$ is the signal propagating time $\frac{d_{ij,n}+s_{ij,n}}{c}$ with the Euclidean distance $d_{ij,n}=\lVert\bm{x}_{i,n}-\bm{x}_{j,n}\rVert$, $s_{ij,n}$ the measurement bias caused by NLOS propagation and speed of the light $c$. Time division multiple access (TDMA) scheme is employed to avoid the collision of packets \cite{sathyan2011wasp}.

The time stamps that the nodes record are the local clock readings $c_{i}(t^{\mathcal T}_{i\to j, n})$ and $c_{j}(t^{\mathcal R}_{i\to j,n})$.
Hence, the observed signal propagation time can be obtained from time stamps $c(t)$ as follows
\begin{align}\label{observation}
 t_{ij,n}=c_{j}(t^{\mathcal R}_{i\to j,n})&-c_{i}(t^{\mathcal T}_{i\to j, n})+\omega_{ij,n}=\frac{\lVert\bm{x}_{i,n}-\bm{x}_{j,n}\rVert+s_{ij,n}}{c}+(\theta_{j,n}-\theta_{i,n})+\omega_{ij,n},
\end{align}
where $\omega_{ij,n}$ is assumed to be Gaussian distributed, ${\omega _{ij,n}} \sim {\cal N}\left(0,\sigma _t^2\right)$.
Multiplying both sides of \eqref{observation} by $c$ we have
\begin{align}\label{rangeobservation}
 z_{ij,n}=\lVert\bm{x}_{i,n}-\bm{x}_{j,n}\rVert+s_{ij,n}+c\,(\theta_{j,n}-\theta_{i,n})+\zeta_{ij,n}\, ,
\end{align}
where $\zeta_{ij,n}=c\cdot\omega_{ij,n}$ is also Gaussian distributed, ${\zeta _{ij,n}} \sim {\cal N}\left(0,c^2\sigma _t^2\right)$. For simplicity we denote $\sigma_d^2=c^2\sigma _t^2$. Since NLOS propagation increases the time that a signal travels between two nodes, the bias $s_{ij,n}$ is positive for that condition. In the case of LOS, no bias is added and $s_{ij,n}$ is zero. Therefore, we have
\begin{align}
s_{ij,n}=\left\{
\begin{array}{cc}
0,& ~~\textrm{if~$(i,j)\notin \Omega_n$,}\\
b_{ij,n},& ~~\textrm{if~$(i,j)\in \Omega_n$.}
\end{array}
\right.
\end{align}s
where $\Omega_n$ denotes the set which contains the pairwise node $(i,j)$ if and only if the measurement $z_{ij,n}$ is NLOS at time $n$. As in \cite{gezici2004uwb}, we model $b_{ij,n}$ as exponential distributed random variable, $b_{ij,n} \sim p(b_{ij,n})$, with
\begin{align}\label{bias}
p(b_{ij,n})=\lambda e^{-\lambda b_{ij,n}},~~~b_{ij,n}>0,
\end{align}
and $\lambda$ is the parameter rate.\footnote{We assume that the range measurement received by a node has been identified to be LOS or NLOS, which can be performed using NLOS identification methods in \cite{wymeersch2012machine}.}

Define $\bm{{{x}}}_n\triangleq[\bm{x}^T_{1,n},\bm{x}^T_{2,n},...,\bm{x}^T_{A+M,n}]^T$ the location variables of all agent and anchor nodes, $\bm{\theta}_n\triangleq[\theta_{1,n},...\theta_{M,n}]^T$ the clock offsets of all agent nodes,  $\bm{{{z}}}_n\triangleq[...,z_{ij,n},...]^T$, $(i,j)\in\Xi,$ as range measurements between all connected nodes, \rd{$\bm{{{b}}}_n\triangleq[...,b_{ij,n},...]^T$, $(i,j)\in\Omega_n,$ as all the NLOS bias in the wireless network at time $n$.} Furthermore $\mathcal{X}_{1:n}\triangleq\{\bm{x}_1,...\bm{x}_N\}$, $\bm{\Theta}_{1:n}\triangleq \{\bm{\theta}_1,...,\bm{\theta}_n\}$, \rd{$\mathcal{B}_{1:n}\triangleq\{\bm{b}_1,...\bm{b}_N\}$} and $\mathcal {Z}_{1:n}\triangleq\{\bm{z}_1,...\bm{z}_n\}$. The goal is to estimate the location $\bm{x}_{i,n}$ and clock offset $\theta_{i,n}$, $i\in{\cal M}$, based on the observation $\mathcal {Z}_{1:n}$ and the state transition information.

\section{Joint Cooperative Localization and Synchronization Algorithms}
In this section, two joint Bayesian estimator based on TOA measurements are proposed.
\edit{
In particular, the estimation algorithms enable node $i$ to estimate its location $\bm{x}_{i,n}$ and clock offset $\theta_{i,n}$ at time $n$ according to the minimum mean square error (MMSE) criteria as
\begin{align}
  \label{eq:estimate}
  \hat{\xi}_{i,n} & = \int \xi_{i,n} \, p\left(\xi_{i,n}|\mathcal {Z}_{1:n} \right) \ist \mathrm{d}\xi_{i,n},
\end{align}
where $\xi_{i,n}$ is used as a replacement character for the location coordinates and for the clock offset of node $i$ at time $n$, i.e., $\xi_{i,n} \in \{\bm{x}_{i,n}, \theta_{i,n}\}$, $p\left(\xi_{i,n}| \mathcal {Z}_{1:n}\right)$ is the marginal posterior distribution given the observations.
}
\subsection{Probabilistic Model}
Assume $\bm{x}_{i,n}$ and $\theta_{i,n}$ evolve according to a memoryless Gauss-Markov process, i.e.,
\begin{align}
\bm{x}_{i,n}&=\bm{x}_{i,n-1}+\bm{v}_{i,n}\, \Delta_t+\bm{\alpha}_n,\\
\theta_{i,n}&=\theta_{i,n-1}+\beta_n,
\end{align}
where $\bm{\alpha}_n$ is Gaussian distributed with zero mean and covariance matrix $\bm{\Sigma}_\alpha=\textrm{diag}\{\sigma_{ux,n}^2,\sigma_{uy,n}^2\}$ and $\beta_n$ is also zero mean Gaussian noise with variance $\sigma_{u\theta,n}^2$, $\Delta_t$ is the time interval and $\bm{v}_{i,n}$ is the velocity at time $n$. Since the agent nodes move independently, the state transition function $p(\bm{x}_n|\bm{x}_{n-1})=\prod_{i} p(\bm{x}_{i,n}|\bm{x}_{i,n-1})$ and $p(\mathcal {X}_{1:n})=p(\bm{x}_0)\prod_{n} p(\bm{x}_n|\bm{x}_{n-1})$, $p(\bm{x}_0)$ is denoted as the prior distribution of all the agent nodes' positions at time $0$. The clock offset can be modeled in the same way. \rd{If the measurement between node $i$ and $j$ is NLOS at time $n$, the prior distribution of bias $p(b_{ij,n})$ is given by \eqref{bias}.}

\edit{The marginal posterior distribution in \eqref{eq:estimate} is computed according to}
\edit{
\begin{align}
  p\left(\xi_{i,n}|{\mathcal {Z}_{1:n}} \right)= & \int  p\left(\mathcal {X}_{1:n},\bm{\Theta}_{1:n},\rd{\mathcal {B}_{1:n}}|\mathcal {Z}_{1:n}\right) \ist \sim\rmv\rmv\{ \mathrm{d}\xi_{i,n} \}
\end{align}
where $p\left(\mathcal {X}_{1:n},\bm{\Theta}_{1:n},{\mathcal {\rd{B}}_{1:n}}|\mathcal {Z}_{1:n}\right)$ is the joint \emph{a posteriori} distribution, and ${\sim\rmv\rmv\{ \textrm{d}\xi_{i,n} \}}$ denotes the integration over all variables collected in $\mathcal {X}_{1:n}$ , $\bm{\Theta}_{1:n}$ and ${\mathcal {\rd{B}}_{1:n}}$ except the variable represented by $\xi_{i,n}$.
}
Using Bayesian rule, we have
\begin{align} \label{joint_example}
p\left(\mathcal {X}_{1:n},\bm{\Theta}_{1:n},{\mathcal {\rd{B}}_{1:n}}|\mathcal {Z}_{1:n}\right)
  \propto
  p\left(\mathcal {Z}_{1:n}|\mathcal {X}_{1:n},\bm{\Theta}_{1:n},{\mathcal {\rd{B}}_{1:n}}\right) \, \edit{p\left(\mathcal {X}_{1:n},\bm{\Theta}_{1:n},{\mathcal {\rd{B}}_{1:n}}\right)} \, .
\end{align}


\edit{
Since the range measurements between nodes at different time are conditional independent, we can factorize the likelihood function as
%
\begin{align}
\label{eq:likelihood}
  p\left(\mathcal {Z}_{1:n}|\mathcal {X}_{1:n},\bm{\Theta}_{1:n},{\mathcal {\rd{B}}_{1:n}}\right) = \prod_n \prod_{(i,j)\notin \Omega_n }p^{\textrm{LOS}}_{ij,n} \prod_{(i,j)\in \Omega_n }p^{\textrm{NLOS}}_{ij,n}, \,\,\,{(i,j)\in\Xi},
\end{align}
}
where
\begin{align}\label{likelihood_function}
p^{\textrm{LOS}}_{ij,n}&=\frac{1}{\sqrt{2\pi\sigma_d^2}}\exp\left\{-\frac{\left(z_{ij,n}-\lVert \bm{x}_{j,n}-\bm{x}_{i,n}\rVert-c\,(\theta_{j,n}-\theta_{i,n})\right)^2}{2\sigma_d^2}\right\},\\\label{likelihood_function_nlos}
\rd{p^{\textrm{NLOS}}_{ij,n}}&\rd{=\frac{1}{\sqrt{2\pi\sigma_d^2}}\exp\left\{-\frac{\left(z_{ij,n}-\lVert \bm{x}_{j,n}-\bm{x}_{i,n}\rVert-c\,(\theta_{j,n}-\theta_{i,n})-b_{ij,n}\right)^2}{2\sigma_d^2}\right\}.}
\end{align}

As the location coordinates and clock offsets of nodes are independent, we can rewrite \eqref{joint_example} as
\begin{align}
\label{eq:apriori}
p\left(\mathcal {X}_{1:n},\bm{\Theta}_{1:n},{\mathcal {\rd{B}}_{1:n}}|\mathcal {Z}_{1:n}\right)\propto &\prod_{i\in \mathcal{M}\cup\mathcal {A}} \left[p(\bm{x}_{i,0}) p(\theta_{i,0}) \prod_{n} p(\bm{x}_{i,n}|\bm{x}_{i,n-1}) p(\theta_{i,n}|\theta_{i,n-1})\right]\nonumber\\&\rd{\times \prod_n \prod_{(i,j)\notin \Omega_n }p^{\textrm{LOS}}_{ij,n} \prod_{(i,j)\in \Omega_n }p(b_{ij,n})\,p^{\textrm{NLOS}}_{ij,n}.}
\end{align}
We further assume that the prior distributions $p\left(\bm{x}_{i,0}\right)$ and $p\left(\theta_{i,0}\right)$ are Gaussian. For anchor node $i\in{\cal A}$ without location and timing uncertainties, the prior distributions are Dirac delta function, which can also be considered as Gaussian distribution with variance equals to zero. The state transition function of anchor node are $p(\bm{x}_n|\bm{x}_{n-1})=\delta(\bm{x}_n-\bm{x}_{n-1})$ and $p(\theta_n|\theta_{n-1})=\delta(\theta_n-\theta_{n-1})=\delta(\theta_n)$.

%


\subsection{Factor Graph Representation}

\edit{We aim to depict the \emph{a posteriori} distribution in \eqref{joint_example} with the factorization \eqref{eq:apriori} by means of a FG \cite{loeliger2004introduction}.}
The
\edit{FG}
is a way to graphically show the mathematical relation between variables and factors. In a FG, there is a factor vertex, drawn as rectangular for every local function and a variable vertex, drawn as circle, for every variable. The factor vertex is connected with a variable vertex if and only if the factor is a function of this variable.



\begin{figure}
\centering
\includegraphics[width=.65\textwidth]{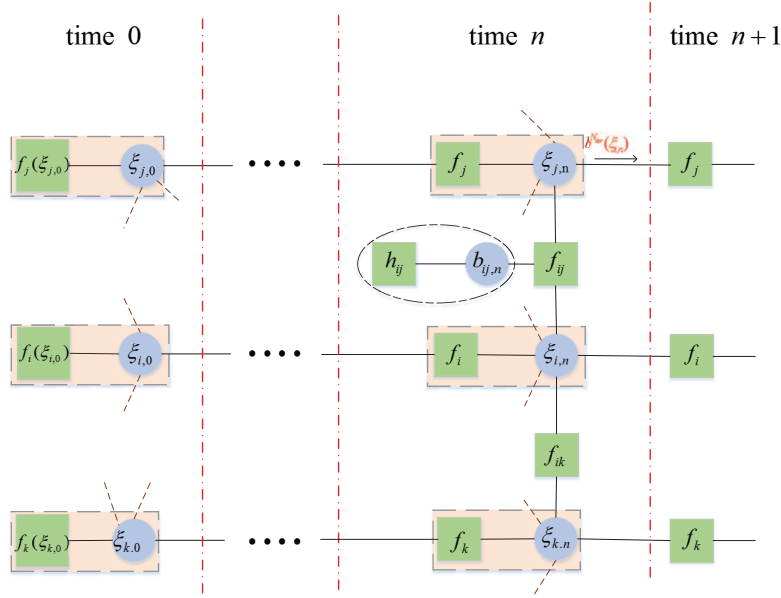}
\caption{\quad A plate model based factor graph for joint localization and synchronization problem. For brevity, the variable node $\xi_{i,n}$ denotes the position or time offset. \rd{The dashed ellipse indicates that at time $n$ the measurement between node $i$ and $j$ is NLOS, which means the corresponding likelihood function $f_{ij}=p^{\textrm{NLOS}}_{ij,n}$.}}
\label{Fig33}
\centering
\end{figure}

\edit{Using the following simplified notation at time $n$}
\begin{align*}
    &f_i\left({\xi}_{i,n}|{\xi}_{i,n-1}\right)=p\left(\xi_{i,n}|{\xi}_{i,n-1}\right),\nonumber\\
    &h_{ij}=p(b_{ij,n}),\nonumber\\
    &f_{ij}=\left\{
    \begin{array}{cc}
    p^{\textrm{LOS}}_{ij,n}&~~{(i,j)\notin \Omega_n}\\
    p^{\textrm{NLOS}}_{ij,n}&~~{(i,j)\in \Omega_n}\, ,
    \end{array}
    \right.
\end{align*}
and the concept of plate models \cite{buntine1994operations}, the joint \emph{a posteriori} distribution in \eqref{eq:apriori} is represented by the FG in Fig.~\ref{Fig33}. Without loss of generality, at time $0$, $f_i (\xi_{i,0})=p(\xi_{i,0})$. Note that every plate corresponds to a node $i \in \mathcal{M}\cup\mathcal{A}$.


In the subsequent sections we will perform message passing on FG in Fig.~\ref{Fig33} in order to obtain beliefs $b(\xi_{i,n})$ which approximates the marginals $p(\xi_{i,n}| \mathcal {Z}_{1:n})$ in \eqref{eq:estimate}, i.e., $b(\xi_{i,n})\simeq p\left(\xi_{i,n}| \mathcal {Z}_{1:n}\right)$. Two message passing rules, namely BP and VMP, are going to be used to derive the expressions of messages\footnote{Messages on FG in Fig.~\ref{Fig33} flow only forward in time, since the network connectivity may have changed and the state information would be outdated.}.

\subsection{Belief Propagation-based Algorithm}\label{BP}
There are two kinds of messages in
\edit{BP},
the message from factor vertex to variable vertex and message from variable vertex to factor vertex. The message from factor vertex $f_i$ to variable vertex $\xi_{i,n}$ at time $n$ is
\begin{align}\label{f_itoxi}
\mu_{f_i\to \xi_{i,n}}(\xi_{i,n})=\int f_i({\xi}_{i,n}|{\xi}_{i,n-1})b^{N_{iter}}({\xi}_{i,n-1})\textrm{d}{\xi}_{i,n-1},
\end{align}
the superscript of $b$ denotes the message passing iteration {and the number of iterations is set to $N_{iter}$ at time $n-1$}.
The messages from factor vertex
\edit{$f_{\edit{ij}}$}
to variable vertex $\xi_{\edit{i,n}}$ at the $l$-th message passing iteration at time $n$ is given by
\begin{align}\label{f_to_xi}
\mu_{f_{\edit{ij}}\to {\xi}_{i,n}}^{(l)}\left({\xi}_{i,n}\right)&= \idotsint \rmv \rmv f_{ij} \prod_{\vartheta\in \mathcal F_{ij,n}\slash \xi_{i,n}} \hspace{-3mm}  \mu^{(l-1)}_{\vartheta\to f_{\edit{ij}}} \rmv (\vartheta) \,\mathrm{d} \vartheta  ,
\end{align}
with $\mu^{(l-1)}_{{\vartheta}\to f_{\edit{ij}}}\left({\vartheta}\right)$ the message from variable vertex $\vartheta$ to $f_{ij}$ at the $(l-1)$-th iteration. $\mathcal F_{\edit{ij},n}$ denotes the set of all variable vertices connected with factor vertex $f_{\edit{ij}}$. The message from variable vertices to factor vertex $f_{\edit{ij}}$ can be updated as
\begin{align}\label{x_to_f}
    \mu^{(l)}_{{\xi}_{i,n}\to f_{\edit{ij}}}\left({\xi}_{i,n}\right)&=\mu_{f_i\to \xi_{i,n}} \prod_{j^{'}\in {\cal S}_{i,n}\slash j}\mu^{(l)}_{f_{ij^{'}}\to \xi_{i,n}}\left({\xi}_{i,n}\right),\\
    \rd{\mu^{(l)}_{{b}_{ij,n}\to f_{{ij}}}\left({b}_{ij,n}\right)}&\rd{=\mu_{h_{ij}\to {b}_{ij,n}}\left({b}_{ij,n}\right)=p\left({b}_{ij,n}\right),}
\end{align}
where ${\cal S}_{i,n}\slash j$ is the set of all neighboring nodes of $\xi_{i,n}$ except node $j$.

After obtaining all the messages by \eqref{f_itoxi} and \eqref{f_to_xi} directed to variable vertex $\xi_{i,n}$, the belief of variable $\xi_{i,n}$ at the $l$-th iteration can be calculated by
\begin{align}\label{postx}
b^{(l)}_{{\xi}_{i,n}}\left({\xi}_{i,n}\right)&=\mu_{f_i\to \xi_{i,n}} (\xi_{i,n}) \prod_{j\in {\cal S}_{i,n}}\mu^{(l)}_{f_{ij}\to \xi_{i,n}}\left({\xi}_{i,n}\right).
\end{align}
Then
\edit{in accordance to \eqref{eq:estimate}},
the location coordinates and the clock offset can be \edit{approximately} determined by
\begin{align}\label{MMSE}
  \hat{\xi}_{i,n} 
    \simeq\int{\xi_{i,n}} \, b^{(N_{iter})}_{{\xi}_{i,n}}\left(\xi_{i,n}\right) \, \mathrm{d}\xi_{i,n} \,.
\end{align}
The bias can also be estimated in a similar way, details are not given in this paper for space limitation. \edit{Note that all the messages related to a variable vertex $\xi_{i,n}$ are locally computed at node $i$, and that the messages $\mu_{\xi_{i,n} \to f_{ij}}^{(l)}(\xi_{i,n})$ are transmitted to the corresponding neighbor $j \in \mathcal{M}_{i,n}$. Hence, the belief in \eqref{postx} and the estimate in \eqref{MMSE} can be obtained by local computations at node $i$ only.}

\edit{We will now consider the computation of \eqref{f_to_xi} in detail.}
{Expanding the exponent in \eqref{likelihood_function_nlos} yields\footnote{Note that \eqref{likelihood_function} can be expanded in a similar way. The results can be obtained by removing the term related to bias $b_{ij,n}$.}
\rd{
\begin{align}\label{exponent}
p^{\textrm{NLOS}}_{ij,n} = \frac{1}{\sqrt{2 \pi \sigma_d^2}} \exp\Bigg\{-&\frac{e_{ij,n} + \varepsilon_{ij,n}}{\sigma_d^2} \Bigg\},
\end{align}
with
\begin{align}
  e_{ij,n} =& z_{ij,n}^2 +(x_{j,n}-x_{i,n})^2+(y_{j,n}-y_{i,n})^2 + c^2 \ist (\theta_{j,n}-\theta_{i,n})^2 + 2 z_{ij,n}c \ist (\theta_{j,n}-\theta_{i,n})\nonumber\\&+ b_{ij,n}^2+2 z_{ij,n} b_{ij,n}+ 2c \ist (\theta_{j,n}-\theta_{i,n}) b_{ij,n}\, , \\\label{squarerootterm}
  \varepsilon_{ij,n} =&-2\left[z_{ij,n}+c\,(\theta_{j,n}-\theta_{i,n})+b_{ij,n} \right]\sqrt{(x_{j,n}-x_{i,n})^2+(y_{j,n}-y_{i,n})^2} \, .
\end{align}
}

Due to the {square root term in \eqref{squarerootterm}}, even if the message from variable vertex $\xi_{i,n}$ to factor vertex $f_{\edit{ij}}$ at the $(l-1)$-th iteration is Gaussian, i.e.,
\begin{align}
  \mu^{(l-1)}_{{\xi}_{i,n}\to f_{ij}}\left({\xi}_{i,n}\right)={\cal N}\left(\xi_{i,n},m^{(l-1)}_{\xi_{i,n}\to f_{ij}},\left(\sigma^{(l-1)}_{\xi_{i,n}\to f_{ij}}\right)^2\right) \, ,\nonumber
\end{align}
the evaluation of \eqref{f_to_xi} \edit{in closed form} is still intractable. Particle-based method can solve this problem but suffers high communication cost.
To this end, we use the first order of Taylor expansion to linearize the square root term in the exponent of the likelihood function. Accordingly, all the messages on this FG can be expressed in closed-form Gaussian expressions.

At the $l$-th iteration, we expand the square root terms in
\edit{$\varepsilon_{ij,n}$}
according to Taylor series around node $i$'s and node $j$'s location estimations $(\hat{x}_{i,n}^{(l-1)},\hat{y}_{i,n}^{(l-1)})$ and $(\hat{x}_{j,n}^{(l-1)},\hat{y}_{j,n}^{(l-1)})$ at the $(l-1)$-th iteration, i.e.,

\begin{align}\label{sqrt_approximate}
    \sqrt{\left(x_{i,n}-x_{j,n}\right)^2+\left(y_{i,n}-y_{j,n}\right)^2}\simeq&
     \hat{d}_{ij,n}^{(l-1)}+\lambda^{\edit{(l-1)}}_{\edit{ij,}n} \left(x_{i,n}-\hat{x}_{i,n}^{(l-1)}\right) +\gamma^{\edit{(l-1)}}_{\edit{ij,}n} \left(y_{i,n}-\hat{y}_{i,n}^{(l-1)}\right) \nonumber\\
    &+\lambda^{\edit{(l-1)}}_{\edit{ij,}n}\left(\hat{x}_{j,n}^{(l-1)}-x_{j,n}\right)
    +\gamma^{\edit{(l-1)}}_{\edit{ij,}n}\left(\hat{y}_{j,n}^{(l-1)}-y_{j,n}\right),
\end{align}
where $\hat{d}_{ij,n}^{(l-1)}\triangleq\sqrt{\left(\hat{x}_{j,n}^{(l-1)}-\hat{x}_{i,n}^{(l-1)}\right)^2+\left(\hat{y}_{j,n}^{(l-1)}-\hat{y}_{i,n}^{(l-1)}\right)^2}$ is the rang\edit{e} estimat\edit{e} obtained in the $(l-1)$-th iteration. $\lambda^{\edit{(l-1)}}_{\edit{ij,}n}=\frac{\hat{x}_{i,n}^{(l-1)}-\hat{x}_{j,n}^{(l-1)}}{\hat{d}_{ij,n}^{(l-1)}}$ and $\gamma^{\edit{(l-1)}}_{\edit{ij,}n}=\frac{\hat{y}_{i,n}^{(l-1)}-\hat{y}_{j,n}^{(l-1)}}{\hat{d}_{ij,n }^{(l-1)}}$ are the directional derivatives on $x$-axis and $y$-axis.

\begin{figure}
\centering
\includegraphics[width=.6\textwidth]{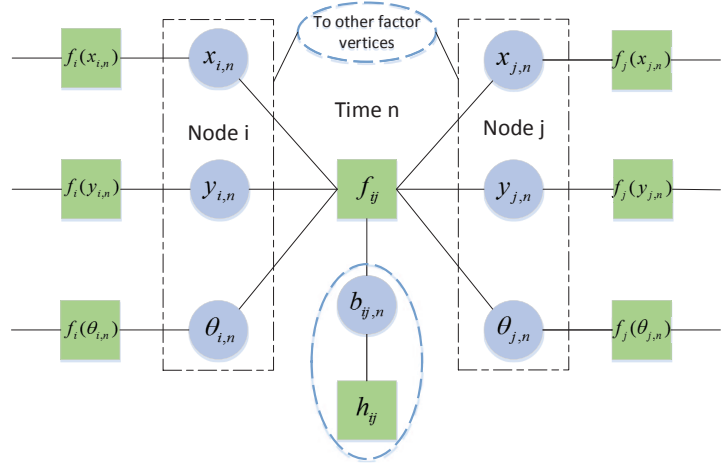}
\caption{Factor graph of a node pair $(i,j)\in\Xi$ at time $n$. The dashed eclipse contains the bias and its corresponding prior, which indicates $(i,j)\in \Omega_n$. If $z_{ij,n}$ is LOS measurement, the dashed eclipse can be removed.}
\label{Fig4}
\centering
\end{figure}

Specifically, if $j\in{\cal A}$, the position of anchor node $j$ is accurately known, the Taylor expansion only operates around node $i$'s previous estimated position. The $(l-1)$-th estimation $(\hat{x}_{j,n}^{(l-1)},\hat{y}_{j,n}^{(l-1)})$ of node $j$ is replaced by its true position $(m_{x_{j,n}},m_{y_{j,n}})$, the directional derivatives become
$\lambda^{\edit{(l-1)}}_{\edit{ij,}n}=\frac{\hat{x}_{i,n}^{(l-1)}-m_{{x}_{j,n}}}{\hat{d}_{ij,n}^{(l-1)}}$ and $\gamma^{\edit{(l-1)}}_{\edit{ij,}n}=\frac{\hat{y}_{i,n}^{(l-1)}-m_{{y}_{j,n}}}{\hat{d}_{ij,n}^{(l-1)}}$. With the first order Taylor expansion, the coordinates of $x$-axis and $y$-axis are conditionally independent given the clock offsets. Therefore the messages to $x_{i,n}$ and $y_{i,n}$ can be calculated separately and hold $\mu_{f\to \bm{x}_{i,n}}(\bm{x}_{i,n})=\mu_{f\to x_{i,n}}(x_{i,n})\, \mu_{f\to y_{i,n}}(y_{i,n})$. The independent assumption of $x$-axis and $y$-axis has also been considered in \cite{chen2006network}. Without loss of generality, here we redefine the variable $\xi_{i,n}\in\{x_{i,n},y_{i,n},\theta_{i,n}\}$. Based on the above linearization, the plate representation of FG in Fig.~\ref{Fig33} can be resolved in Fig.~\ref{Fig4}, where a single connection $(i,j)\in \Xi$ is depicted.

\edit{
Using the approximation \eqref{sqrt_approximate} in the LOS likelihood function \eqref{likelihood_function}, we are now able to express the messages $\mu^{(l)}_{f_{ij} \to \xi_{i,n}}$ in Gaussian form, i.e.,
\begin{align}\label{factortovariable}
  \mu^{(l)}_{f_{ij} \to \xi_{i,n}}(\xi_{i,n}) \propto \mathcal{N}\left(\xi_{i,n},m^{(l)}_{f_{ij} \to \xi_{i,n}},\left(\sigma^{(l)}_{f_{ij} \to \xi_{i,n}}\right)^2 \right) \, .
\end{align}
}
If node $j$ is an anchor node,
\edit{
we have the parameters for the location coordinates as
\begin{align}
  m^{(l)}_{f_{ij} \to x_{i,n}} & = m_{x_{j,n}}+\left(z_{ij,n}-c \ist m^{(l-1)}_{{\theta_{i,n}}\to {f_{ij}}}\right) \ist \lambda^{(l-1)}_{ij,n}, \label{BPftox} \\
  m^{(l)}_{f_{ij} \to y_{i,n}} & = m_{y_{j,n}}+\left(z_{ij,n}-c \ist m^{(l-1)}_{{\theta_{i,n}}\to {f_{ij}}}\right) \ist \gamma^{(l-1)}_{ij,n}, \label{BPftoy}\\
  \left(\sigma^{(l)}_{f_{ij} \to x_{i,n}}\right)^2 & = \left(\sigma^{(l)}_{f_{ij} \to y_{i,n}}\right)^2 = {c^2} \left(\sigma^{(l-1)}_{{\theta_{i,n}}\to f_{ij}}\right)^2+\sigma^2_d,
\end{align}
and the parameters for the clock offset as
\begin{align}\label{BPftothetamean}
 m^{(l)}_{f_{ij} \to \theta_{i,n}} \!= & \frac{z_{ij,n}}{c} \!-\! \frac{1}{c^2\!\left(\sigma_{xyd,n}^{(l-1)}\right)^2}\! \left[ \left(\sigma_{xd,n}^{(l-1)}\right)^2 \!\!\left(\sigma_{yd,n}^{(l-1)}\right)^2 \! \hat{d}_{ij,n}^{(l-1)} \!-\! \left(\sigma_{xd,n}^{(l-1)}\right)^2\! \left(\sigma^{(l-1)}_{{y_{i,n}}\to {f_{ij}}}\right)^2 \!\gamma^{(l-1)}_{ij,n} \! \left(m^{(l-1)}_{{x_{i,n}}\to {f_{ij}}}\!-\!m_{x_{{j,n}}}\right) \right. \nonumber\\
 & \ \ \ \ \ \ \ \ \ \ \ \ \ \ \ \ \ \ \ \ \ \ \ \ \left. + \left(\sigma_{yd,n}^{(l-1)}\right)^2 \left(\sigma^{(l-1)}_{{x_{i,n}}\to {f_{ij}}}\right)^2 \lambda^{(l-1)}_{ij,n} \left(m^{(l-1)}_{{y_{i,n}}\to {f_{ij}}}-m_{y_{{j,n}}}\right)\right],\\ \label{BPftothetavar}
 &\hspace{3cm}\left(\sigma^{(l)}_{f_{ij} \to \theta_{i,n}}\right)^2 =\frac{\left(\sigma_{xd,n}^{(l-1)}\right)^2 \left(\sigma_{yd,n}^{(l-1)}\right)^2}{c^2\left(\sigma_{xyd,n}^{(l-1)}\right)^2},
\end{align}
with $(\sigma_{xd,n}^{(l-1)})^2 \triangleq (\sigma^{(l-1)}_{{x_{i,n}}\to {f_{ij}}})^2+{\sigma^2_d}$, $(\sigma_{yd,n}^{(l-1)})^2\triangleq (\sigma^{(l-1)}_{{y_{i,n}}\to {f_{ij}}})^2+{\sigma^2_d}$ and $(\sigma_{xyd,n}^{(l-1)})^2 \triangleq \sigma^2_d+(\sigma^{(l-1)}_{{x_{i,n}}\to {f_{ij}}} {\lambda^{(l-1)}_{ij,n}})^2 +(\sigma^{(l-1)}_{{y_{i,n}}\to {f_{ij}}}{\gamma^{(l-1)}_{ij,n}})^2$.
}

If node $j$ is an agent node, the parameters for location coordinates are determined as
\begin{align}
 m^{(l)}_{f_{ij} \to x_{i,n}}&= m^{(l-1)}_{{x_{j,n}}\to f_{ij}}+\left(z_{ij,n}-c\,\left(m^{(l-1)}_{\theta_{{i,n}}\to f_{ij}}-m^{(l-1)}_{\theta_{{j,n}}\to f_{ij}}\right)\right)\, \lambda^{\edit{(l-1)}}_{\edit{ij,}n},\\
  m^{(l)}_{f_{ij} \to y_{i,n}}&= m^{(l-1)}_{{y_{j,n}}\to f_{ij}}+\left(z_{ij,n}-c\,\left(m^{(l-1)}_{\theta_{{i,n}}\to f_{ij}}-m^{(l-1)}_{\theta_{j,n}\to f_{ij}}\right)\right)\, \gamma^{\edit{(l-1)}}_{\edit{ij,}n},\\
  \left(\sigma^{(l)}_{f_{ij}\to x_{i,n}}\right)^2 &={c^2}\left(\sigma^{(l-1)}_{{\theta_{i,n}}\to {f_{ij}}}\right)^2+{c^2}\left(\sigma^{(l-1)}_{{\theta_{j,n}}\to {f_{ij}}}\right)^2+\left(\sigma^{(l-1)}_{{x_{j,n}}\to {f_{ij}}}\right)^2+\sigma^2_d,\\
    \left(\sigma^{(l)}_{f_{ij}\to y_{i,n}}\right)^2 &={c^2}\left(\sigma^{(l-1)}_{{\theta_{i,n}}\to {f_{ij}}}\right)^2+{c^2}\left(\sigma^{(l-1)}_{{\theta_{j,n}}\to {f_{ij}}}\right)^2+\left(\sigma^{(l-1)}_{{y_{j,n}}\to {f_{ij}}}\right)^2+\sigma^2_d,
\end{align}
and the parameters for the clock offset as
\begin{align}\label{BPagenttothetamean}
 m^{(l)}_{f_{ij} \to \theta_{i,n}} \!=&  \frac{z_{ij,n}}{c} - \frac{1}{c^2\left(\sigma_{xy,n}^{(l-1)}\right)^2} \left[ \left(\sigma_{x,n}^{(l-1)}\right)^2 \! \left(\sigma_{y,n}^{(l-1)}\right)^2 \! \hat{d}_{ij,n}^{(l-1)} \right. \nonumber \\
 & \left. -\left(\sigma_{x,n}^{(l-1)}\right)^{2} \left((\sigma^{(l-1)}_{{y_{i,n}}\to {f_{ij}}})^2+(\sigma^{(l-1)}_{{y_{j,n}}\to {f_{ij}}})^2\right)\gamma^{(l-1)}_{ij,n} \left(m^{(l-1)}_{{x_{i,n}}\to {f_{ij}}}-m^{(l-1)}_{{x_{j,n}}\to {f_{ij}}}\right)  \right. \nonumber\\
& \left. + \left(\sigma_{y,n}^{(l-1)}\right)^{2} \left((\sigma^{(l-1)}_{{x_{i,n}}\to {f_{ij}}})^2+(\sigma^{(l-1)}_{{x_{j,n}}\to {f_{ij}}})^2\right) \lambda^{(l-1)}_{ij,n} \left(m^{(l-1)}_{{y_{i,n}}\to {f_{ij}}}-m^{(l-1)}_{{y_{j,n}}\to {f_{ij}}}\right) \right],
 \\\label{BPagenttothetavar}
&\hspace{3cm} \left(\sigma^{(l)}_{f_{ij} \to \theta_{i,n}}\right)^2 =  \frac{\left(\sigma_{x,n}^{(l-1)}\right)^2 \left(\sigma_{y,n}^{(l-1)}\right)^2}{c^2\left(\sigma_{xy,n}^{(l-1)}\right)^2},
\end{align}
where $\lambda^{\edit{(l-1)}}_{\edit{ij,}n},\,\gamma^{\edit{(l-1)}}_{\edit{ij,}n}$ are the directional derivatives, $(\sigma_{x,n}^{(l-1)})^{2}\!\triangleq\!(\sigma^{(l-1)}_{{x_{i,n}}\to {f_{ij}}})^2+(\sigma^{(l-1)}_{{x_{j,n}}\to {f_{ij}}})^2+\sigma_d^2$, $(\sigma_{y,n}^{(l-1)})^{2}\triangleq(\sigma^{(l-1)}_{{y_{i,n}}\to {f_{ij}}})^2+(\sigma^{(l-1)}_{{y_{j,n}}\to {f_{ij}}})^2+\sigma_d^2$ and $(\sigma_{xy,n}^{(l-1)})^2\triangleq\sigma_d^2+(\lambda^{\edit{(l-1)}}_{\edit{ij,}n})^2 \big((\sigma^{(l-1)}_{{x_{i,n}}\to {f_{ij}}})^2+(\sigma^{(l-1)}_{{x_{j,n}}\to {f_{ij}}})^2\big)+(\gamma^{\edit{(l-1)}}_{\edit{ij,}n})^2\big((\sigma^{(l-1)}_{{y_{i,n}}\to {f_{ij}}})^2+(\sigma^{(l-1)}_{{y_{j,n}}\to {f_{ij}}})^2\big)$.

\rd{For NLOS measurement, since the bias is exponentially distributed, the integrals in \eqref{f_to_xi} can not be expressed in Gaussian closed form. We approximate \eqref{f_to_xi} to a Gaussian message by moment-matching, i.e.,
\begin{align}\label{moment}
m^{(l)}_{f_{ij} \to \xi_{i,n}}&=\mathbb{E}_{\xi_{i,n}}[{\mu}^{(l)}_{f_{ij} \to \xi_{i,n}}],\\\label{moment1}
\left(\sigma^{(l)}_{f_{ij} \to \xi_{i,n}}\right)^2&=\mathbb{E}_{\xi_{i,n}}[({\mu}^{(l)}_{f_{ij} \to \xi_{i,n}})^2]-\mathbb{E}^2_{\xi_{i,n}}[{\mu}^{(l)}_{f_{ij} \to \xi_{i,n}}].
  \end{align}
}

The messages from $f_i$ can also be determined with respect to $x$, $y$ and $\theta$ as
\begin{align}\label{f_itox_i}
\mu_{f_i\to x_{i,n}}(x_{i,n})&=\mathcal {N} \left(x_{i,n},m^{(N_{iter})}_{x_{i,n-1}}+v_{x_{i,n-1}}\,\Delta_t,\left(\sigma^{(N_{iter})}_{x_{i,n-1}}\right)^2+\sigma^{2}_{ux,n}\right),\\
\mu_{f_i\to y_{i,n}}(y_{i,n})&=\mathcal {N} \left(y_{i,n},m^{(N_{iter})}_{y_{i,n-1}}+v_{y_{i,n-1}}\,\Delta_t,\left(\sigma^{(N_{iter})}_{y_{i,n-1}}\right)^2+\sigma^{2}_{uy,n}\right),\\\label{f_itotheta_i}
\mu_{f_i\to \theta_{i,n}}(\theta_{i,n})&=\mathcal {N} \left(x_{i,n},m^{(N_{iter})}_{\theta_{i,n-1}},\left(\sigma^{(N_{iter})}_{\theta_{i,n-1}}\right)^2+\sigma_{u\theta,n}^2\right).
\end{align}
After collecting all the incoming messages from connected factor vertices, we can calculate the belief{ of variable $\xi_{i,n}$ }by using \eqref{postx}. Since all the incoming messages are Gaussian, the product of multiple Gaussian distributions can also be written in Gaussian form \cite{aldershof1995facts}, which gives
\begin{align}\label{result_x_marginal}
    b^{(l)}\left(\xi_{i,n}\right)&\propto{\cal N}\left(\xi_{i,n},m_{\xi_{i,n}}^{(l)},\left(\sigma_{\xi_{i,n}}^{(l)}\right)^2\right),
\end{align}
where the mean and variance in \eqref{result_x_marginal} are
\begin{align}
 {m}^{(l)}_{\xi_{i,n}}=\left({\sigma}^{(l)}_{\xi_{i,n}}\right)^2\left(\frac{m_{f_i\to \xi_{i,n}}}{\sigma_{f_i\to \xi_{i,n}}^2}+\sum_{j\in{{\cal S}_{i,n}}}\frac{m^{(l)}_{f_{ij}\to \xi_{i,n}}}{\left(\sigma^{(l)}_{f_{ij}\to \xi_{i,n}}\right)^2}\right),\label{mx}\\
\left({\sigma}^{(l)}_{\xi_{i,n}}\right)^2=\left(\left(\sigma_{f_i\to \xi_{i,n}}\right)^{-2}+\sum_{j\in{\cal S}_{i,n}}\left(\sigma^{(l)}_{f_{ij}\to \xi_{i,n}}\right)^{-2}\right)^{-1}.\label{sigmax}
\end{align}


Then, the message from variable vertex $\xi_{i,{n}}$ to a connected factor vertex $f_{ij}$, $j\in \mathcal {M}_{i,n}$, can be calculated as Gaussian distribution $\mathcal{N}\left(\xi_{i,n}, m^{(l)}_{\xi_{i,n}\to f_{ij}}, \left(\sigma^{(l)}_{\xi_{i,n}\to f_{ij}}\right)^2\right)$, whose mean and variance are
\begin{align}\label{meanxitof}
m^{(l)}_{\xi_{i,n}\to f_{ij}}&=\frac{{ m}^{(l)}_{\xi_{i,n}}\left(\sigma^{(l)}_{f_{ij}\to \xi_{i,n}}\right)^2-m^{(l)}_{f_{ij}\to \xi_{i,n}}\left(\sigma^{(l)}_{\xi_{i,n}}\right)^2}{\left(\sigma^{(l)}_{f_{ij}\to \xi_{i,n}}\right)^2-\left(\sigma^{(l)}_{\xi_{i,n}}\right)^2},\\\label{varxitof}
\left(\sigma^{(l)}_{\xi_{i,n}\to f_{ij}}\right)^2&=\frac{\left(\sigma^{(l)}_{\xi_{i,n}}\right)^2\left(\sigma^{(l)}_{f_{ij}\to \xi_{i,n}}\right)^2}{\left(\sigma^{(l)}_{f_{ij}\to \xi_{i,n}}\right)^2-\left(\sigma^{(l)}_{\xi_{i,n}}\right)^2}.
\end{align}

It can be observed that the messages from a variable vertex vary for different factor vertices connecting to it. Calculating all different outgoing messages separately results in high computational complexity. The message from a variable vertex $\xi_{i,n}$ to a factor vertex $f$ is the belief of $\xi_{i,n}$ divided by the message from the factor vertex $f$. That is to say all the messages from the same variable vertex to different factor vertices only differ in one term. When the connectivity of wireless network is high, the difference of these messages are negligible. Based on this observation, we can approximate the messages from a variable vertex to the connected factor vertices by its belief. As a result, a broadcast message passing scheme can be applied. We name this approximated algorithm as \emph{broadcast} BP, while the algorithm following the exact sum-product rules is named as \emph{standard} BP.

The proposed belief propagation algorithm for joint cooperative localization and synchronization is depicted in \textbf{Algorithm} \ref{algorithm1}. Since all the beliefs and messages on the FG are expressed in Gaussian closed form, only the means and variances have to be updated and transmitted, which significantly reduces the computational complexity and the communication overhead.

  \begin{algorithm}
 \allowdisplaybreaks
     \caption{Belief propagation-based joint estimation algorithm}
     \label{algorithm1}
     \begin{algorithmic}[1]
     \STATE {At time $n=0$ \bf{Initialization}}:
         \STATE {nodes} $i\in{\cal A}$ \emph{initialize}
             \STATE ~~~$f_{i}\left(x_{i,0}\right)=\delta\left(x_{i,0}-m_{x_{i,n}}\right)$, $f_{i}\left(y_{i,0}\right)=\delta\left(y_{i,0}-m_{y_{i,n}}\right)$, $f_{i}\left(\theta_{i,0}\right)=\delta\left(\theta_{i,0}\right)$ ;
             \STATE {nodes} $i\in{\cal  M}$ \emph{initialize}
             \STATE~~~$f_{i}\left(x_{i,0}\right)\propto{\cal N}\left(x_{i,0},m_{x_{i,0}},\left(\sigma_{x_{i,0}}\right)^2\right)$, $f_{i}\left(y_{i,0}\right)\propto{\cal N}\left(y_{i,0},m_{y_{i,0}},\left(\sigma_{y_{i,0}}\right)^2\right)$,  \\~~~$f_{i}\left({\theta}_{i,0}\right)\propto{\cal N}\left(\theta_{i,0},m_{\theta_{i,0}},\left(\sigma_{\theta_{i,0}}\right)^2\right)$;\\
  \FOR {$n=1$ to $N_{time}$ (time index)}
           \STATE {nodes} $i\in{\cal M}$ {\bf{in parallel}}
        \STATE ~~~compute the messages from factor vertex $f_i$ to variable vertices $\mu_{{f}_i\to x_{i,n}}(x_{i,n})$, \\~~~$\mu_{{f}_i\to y_{i,n}}(y_{i,n})$ and $\mu_{{f}_i\to \theta_{i,n}}(\theta_{i,n})$ according to \eqref{f_itox_i}-\eqref{f_itotheta_i}.
         \FOR{$l=1$ to $N_{iter}$}
         \STATE ~~\textbf{If} $(i,j)\notin \Omega_n$
                   \STATE  ~~~~~compute all messages from factor vertices to variable vertices $\mu_{f_{ij}\to {\xi}_{i,n}}^{(l)}\left({\xi}_{i,n}\right)$ \\~~~~~according to \eqref{factortovariable};
                   \STATE ~~\textbf{Else} ~~~set $\mu_{b_{ij,n}\to f_{ij}}(b_{ij,n})=\lambda\exp(-\lambda b_{ij,n})$
                    \STATE  ~~~~~compute all messages from factor vertices to variable vertices $\mu_{f_{ij}\to {\xi}_{i,n}}^{(l)}\left({\xi}_{i,n}\right)$ \\~~~~~according to \eqref{moment} and \eqref{moment1};
                   \STATE ~~update the beliefs of location coordinates and clock offset according to \eqref{mx} and \eqref{sigmax};
                   \STATE ~~\emph{Broadcast} BP:
                   \\~~~Broadcast the beliefs to neighboring nodes;
                   \STATE ~~\emph{Standard} BP:
                   \\~~~Compute all messages directed to the connected factor vertices $\mu_{{x}_{i,n}\to f_{ij}}^{(l)}\!\!\left({x}_{i,n}\right)$, \\~~~$\mu_{{y}_{i,n}\!\to\! f_{ij}}^{(l)}\!\!\left({y}_{i,n}\right)$ and $\mu_{{\theta}_{i,n} \to f_{ij}}^{(l)}\!\left({\theta}_{i,n}\right)$ according to \eqref{meanxitof} and \eqref{varxitof} and transmit them to \\~~~neighboring nodes;
         \ENDFOR;
         \STATE transmit the beliefs $b^{(N_{iter})}(x_{i,n})$, $b^{(N_{iter})}(y_{i,n})$ and $b^{(N_{iter})}(\theta_{i,n})$ to the factor vertex $f_i$.
         \STATE estimate agents' positions and clock offsets using {\bf{MMSE}} estimator;
         \STATE {\bf{end parallel}};
         \ENDFOR;
     \end{algorithmic}
 \end{algorithm}
\subsection{Variational Message Passing-based Algorithm}\label{VMP}
Variational methods aim at approximating a complex or intractable distribution by a much simpler one\cite{rockafellar1998variational}. In contrast to BP, VMP imposes that the joint belief fully factorizes. The update rules of VMP on a FG is given in \cite{dauwels2007variational}. Based on the same assumptions in Sec.\,\ref{BP}, incoming messages from factor vertices to variable vertex at the $l$-th iteration are

\begin{align}\label{vmpf_to_xi}
\mu_{f_{ij}\to {\xi}_{i,n}}^{(l)}\left({\xi}_{i,n}\right)&=\exp\bigg(\idotsint\ln {f_{ij}}\prod_{\vartheta\in \mathcal F_{ij,n}\slash \xi_{i,n}}\mu^{(l-1)}_{{\vartheta}\to f_{ij}}\left(\vartheta\right)\,\mathrm{d} \vartheta\bigg),\\
\mu_{f_{ij}\to {b}_{ij,n}}^{(l)}\left({b}_{ij,n}\right)&=\exp\bigg(\idotsint\ln {f_{ij}}\prod_{\vartheta\in \mathcal F_{ij,n}\slash b_{ij,n}}\mu^{(l-1)}_{{\vartheta}\to f_{ij}}\left(\vartheta\right)\,\mathrm{d} \vartheta\bigg), ~~(i,j)\in \Omega_n\\
\mu_{f_{i}\to {\xi}_{i,n}}\left({\xi}_{i,n}\right)&=\exp\left(\int\ln\left({f_{i}(\xi_{i,n}|\xi_{i,n-1})}\right)\, b^{N_{iter}}(\xi_{i,n-1})\,\mathrm{d} \xi_{i,n-1}\right).
\end{align}

\rd{The outgoing messages from variable vertices to the connected factor vertices are obtained as
\begin{align}
    &\mu^{(l)}_{{\xi}_{i,n}\to f_{ij}}\left({\xi}_{i,n}\right)=b^{(l)}\left({\xi}_{i,n}\right)=\prod_{j^{'}\in {\cal S}_{i,n}}\mu^{(l)}_{f_{ij^{'}}\to \xi_{i,n}}\left({\xi}_{i,n}\right),\\
    &\mu^{(l)}_{{b}_{ij,n}\to f_{ij}}\left({b}_{ij,n}\right)=b^{(l)}\left({b}_{ij,n}\right)=\mu^{(l)}_{f_{ij}\to b_{ij,n}}\left(b_{ij,n}\right) \mu_{h_{ij}\to b_{ij,n}}\left(b_{ij,n}\right), ~~(i,j)\in \Omega_n.
\label{vmpx_to_f}
\end{align}}
It can be seen that the messages from variable vertices to the connected factor vertices are the beliefs of variable vertices.
\begin{figure}
\centering
\includegraphics[width=.6\textwidth]{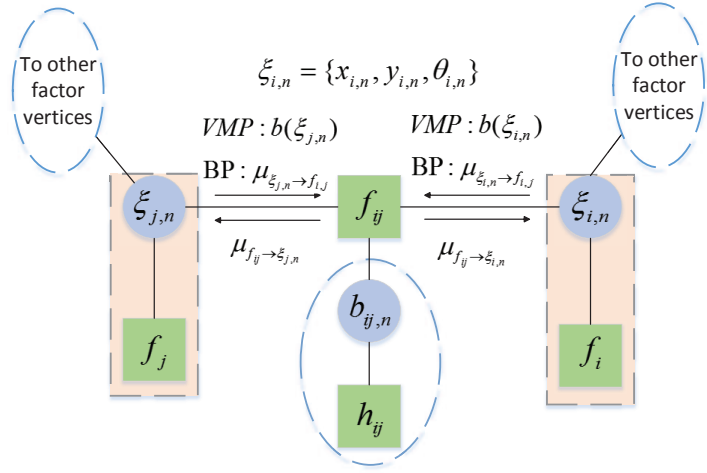}
\caption{\quad Illustration of BP and VMP with plate notation on FG. The dashed eclipse has the same denotation of Fig. 3.}\label{fig5}
\centering
\end{figure}

Note that in VMP, two variable vertices only need to exchange their beliefs instead of the extrinsic information as illustrated in Fig. \ref{fig5}. For a variable vertex, the same information is sent to all the connected factor vertices, hence VMP can be performed in a broadcasting way which significantly reduces communication overhead compared to the \emph{standard} BP.

{As before, we replace the square root in the exponent of \eqref{exponent} by Taylor expansion in \eqref{sqrt_approximate}.} For NLOS measurement, i.e., $(i,j)\in \Omega_n$, if $j$ is an anchor node, the messages from factor vertex to variable vertex are given by
\begin{align}
\label{VMPftox}
&\mu^{(l)}_{f_{ij}\to{x_{i,n}}}(x_{i,n})\propto {\cal N}\left(x_{i,n},m_{x_{j,n}}+\left(z_{ij,n}-c\,\hat\theta^{(l-1)}_{i,n}-\hat{b}^{(l-1)}_{ij.n}\right)\,{\frac{{\hat x}^{(l-1)}_{i,n}-m_{{x}_{j,n}}}{{\hat d}^{(l-1)}_{ij,n}}},\sigma^2_d\right),
\\\label{VMPftoy}
&\mu^{(l)}_{f_{ij}\to{y_{i,n}}}(y_{i,n})\propto {\cal N}\left(y_{i,n},m_{y_{j,n}}+\left(z_{ij,n}-c\,\hat\theta^{(l-1)}_{i,n}-\hat{b}^{(l-1)}_{ij.n}\right)\,{\frac{{\hat y}^{(l-1)}_{i,n}-m_{{y}_{{j,n}}}}{{\hat d}^{(l-1)}_{ij,n}}},\sigma^2_d\right),
\\\label{VMPftotheta}
&\mu^{(l)}_{f_{ij}\to{\theta_{i,n}}}(\theta_{i,n})\propto{\cal N}\left(\theta_{i,n},\frac{z_{ij,n}-{\hat d}^{(l-1)}_{ij,n}-\hat{b}^{(l-1)}_{ij.n}}{c},\frac{\sigma^2_d}{c^2}\right),
\\\label{VMPftobias}
&\mu^{(l)}_{f_{ij}\to{b_{ij,n}}}(b_{ij,n})\propto {\cal N}\left(b_{ij,n},z_{ij,n}-c\,\hat\theta^{(l-1)}_{i,n}-\hat{d}_{ij,n}^{(l-1)},\sigma^2_d\right).
\end{align}
If $j$ is an agent node, the messages becomes
\begin{align}
\label{VMPftox1}
&\hspace{-3mm}\mu^{(l)}_{f_{ij}\to{x_{i,n}}}(x_{i,n})\propto {\cal N}\left(x_{i,n},{\hat{x}^{(l-1)}_{j,n}}+\left(z_{ij,n}-c\,(\hat\theta^{(l-1)}_{i,n}-\hat\theta^{(l-1)}_{j,n})-\hat{b}^{(l-1)}_{ij.n}\right)\,{\frac{{\hat x}^{(l-1)}_{i,n}-\hat{x}^{(l-1)}_{j,n}}{{\hat d}^{(l-1)}_{ij,n}}},\sigma^2_d\right),
\\\label{VMPftoy1}
&\hspace{-3mm}\mu^{(l)}_{f_{ij}\to{y_{i,n}}}(y_{i,n})\propto {\cal N}\left(y_{i,n},{\hat{y}^{(l-1)}_{j,n}}+\left(z_{ij,n}-c\,(\hat\theta^{(l-1)}_{{i,n}}-\hat\theta^{(l-1)}_{j,n})-\hat{b}^{(l-1)}_{ij.n}\right)\,{\frac{{\hat y}^{(l-1)}_{{i,n}}-\hat{y}^{(l-1)}_{j,n}}{{\hat d}^{(l-1)}_{ij,n}}},\sigma^2_d\right),
\\\label{VMPftotheta1}
&\mu^{(l)}_{f_{ij}\to{\theta_{i,n}}}(\theta_{i,n})\propto {\cal N}\left(\theta_{i,n},{\hat{\theta}^{(l-1)}_{j,n}}+\frac{z_{ij,n}-{\hat d}^{(l-1)}_{ij,n}-\hat{b}^{(l-1)}_{ij.n}}{c},\frac{\sigma^2_d}{c^2}\right),
\\\label{VMPftobias1}
&\mu^{(l)}_{f_{ij}\to{b_{ij,n}}}(b_{ij,n})\propto {\cal N}\left(b_{ij,n},z_{ij,n}-c\,(\hat\theta^{(l-1)}_{{i,n}}-\hat\theta^{(l-1)}_{j,n})-\hat{d}_{ij,n}^{(l-1)},\sigma^2_d\right).
\end{align}}
For LOS measurement, i.e., $(i,j)\notin \Omega_n$, the messages can be obtained by removing the term $\hat{b}_{ij,n}$ from \eqref{VMPftox1}-\eqref{VMPftotheta1} straightforwardly.

The messages from $f_i$ to variable vertices are
\begin{align}\label{f_itox_ivmp}
\mu_{f_i\to x_{i,n}}(x_{i,n})&=\mathcal {N} \left(x_{i,n},m^{(N_{iter})}_{x_{i,n-1}}+v_{x_{i,n-1}}\,\Delta_t,\sigma^{2}_{ux,n}\right),\\
\mu_{f_i\to y_{i,n}}(y_{i,n})&=\mathcal {N} \left(y_{i,n},m^{(N_{iter})}_{y_{i,n-1}}+v_{y_{i,n-1}}\,\Delta_t,\sigma^{2}_{uy,n}\right),\\\label{f_itotheta_ivmp}
\mu_{f_i\to \theta_{i,n}}(\theta_{i,n})&=\mathcal {N} \left(x_{i,n},m^{(N_{iter})}_{\theta_{i,n-1}},\sigma_{u\theta,n}^2\right).
\end{align}
It can be seen that the updates of the variances of the messages \eqref{VMPftox}-\eqref{VMPftobias1} in VMP only depend on the variance of measurement noise, while the variances of message \eqref{factortovariable} in BP also rely on the variances of neighboring nodes' positions and clock offsets. This leads to underestimation of variance, which is due to the fact that the mean-field approximation in the proposed VMP assumes that all the variables are independent\cite{hotta2003mean}.

\rd{According to \eqref{VMPftox}-\eqref{VMPftobias1}, the belief of variable $\xi_{i,n}$ and $b_{ij,n}$ can be obtained in Gaussian form, i.e.,
\begin{align}\label{result_x_marginal1}
    &b^{(l)}\left(\xi_{i,n}\right)=\mu^{(l)}_{{\xi_{i,n}}\to f_{ij}}\propto{\cal N}\left(\xi_i,m_{\xi_{i,n}}^{(l)},\left(\sigma_{\xi_{i,n}}^{(l)}\right)^2\right),\\
\label{beliefb1}
&b^{(l)}\left(b_{ij,n}\right)=\mu^{(l)}_{{b_{ij,n}}\to f_{ij}} \propto \mathcal{N}\left(b_{ij,n}, m^{(l-1)}_{f_{ij} \to b_{ij,n}}-\sigma_d^2 \lambda, \sigma_d^2\right),
\end{align}
where the means and variances of \eqref{result_x_marginal1} are}
\begin{align}\label{vmp_mx}
 &{m}^{(l)}_{x_{i,n}}=\left({\sigma}^{(l)}_{x_{i,n}}\right)^2\left(\frac{m_{f_i\to x_{i,n}}}{\sigma_{f_i\to x_{i,n}}^2}+\sum_{j\in{{\cal S}_{i,n}}}\frac{m^{(l)}_{f_{ij}\to x_{i,n}}}{\left(\sigma_d\right)^2}\right),
\end{align}
\begin{align}\label{vmp_my}
 &{m}^{(l)}_{y_{i,n}}=\left({\sigma}^{(l)}_{y_{i,n}}\right)^2\left(\frac{m_{f_i\to y_{i,n}}}{\sigma_{f_i\to y_{i,n}}^2}+\sum_{j\in{{\cal S}_{i,n}}}\frac{m^{(l)}_{f_{ij}\to y_{i,n}}}{\left(\sigma_d\right)^2}\right),\\\label{vmp_mtheta}
 &{m}^{(l)}_{\theta_{i,n}}=\left({\sigma}^{(l)}_{\theta_{i,n}}\right)^2\left(\frac{m_{f_i\to \theta_{i,n}}}{\sigma_{f_i\to \theta_{i,n}}^2}+\sum_{j\in{{\cal S}_{i,n}}}\frac{c^2\cdot m^{(l)}_{f_{ij}\to \theta_{i,n}}}{\left(\sigma_d\right)^2}\right),
\end{align}
\begin{align}
&\left({\sigma}^{(l)}_{x_{i,n}}\right)^2=\left(\left(\sigma_{x_{i,n}}\right)^{-2}+\sum_{j\in{\cal S}_{i,n}}\left(\sigma_d\right)^{-2}\right)^{-1},\label{vmp_sigmax}\\
&\left({\sigma}^{(l)}_{y_{i,n}}\right)^2=\left(\left(\sigma_{y_{i,n}}\right)^{-2}+\sum_{j\in{\cal S}_{i,n}}\left(\sigma_d\right)^{-2}\right)^{-1},\label{vmp_sigmay}\\
&\left({\sigma}^{(l)}_{\theta_{i,n}}\right)^2=\left(\left(\sigma_{\theta_{i,n}}\right)^{-2}+\sum_{j\in{\cal S}_{i,n}}\left(\frac{\sigma_d}{c}\right)^{-2}\right)^{-1}.\label{vmp_sigmatheta}
\end{align}


We have obtained the means and variances of all the messages in VMP-based algorithm in closed form. It is straightforward to estimate the positions and clock offsets of agent nodes using Gaussian beliefs. The proposed VMP algorithm for simultaneous localization and synchronization is described in \textbf{Algorithm} \ref{algorithm2}.
 \begin{algorithm}[h]

     \caption{Variational message passing-based joint estimation algorithm}
     \label{algorithm2}
     \begin{algorithmic}[1]
     \STATE {At time $n=0$ \bf{Initialization}}:
         \STATE {nodes} $i\in{\cal A}$ \emph{initialize}
             \STATE ~~~$f_{i}\left(x_{i,0}\right)=\delta\left(x_{i,0}-m_{x_i}\right)$, $f_{i}\left(y_{i,0}\right)=\delta\left(y_{i,0}-m_{y_i}\right)$, $f_{i}\left(\theta_{i,0}\right)=\delta\left(\theta_{i,0}\right)$ ;
             \STATE {nodes} $i\in{\cal  M}$ \emph{initialize}
             \STATE~~~$f_{i}\left(x_{i,0}\right)\propto{\cal N}\left(x_{i,0},m_{x_{i,0}},\left(\sigma_{x_{i,0}}\right)^2\right)$,
 , $f_{i}\left(y_{i,0}\right)\propto{\cal N}\left(y_{i,0},m_{y_{i,0}},\left(\sigma_{y_{i,0}}\right)^2\right)$,  \\~~~$f_{i}\left({\theta}_{i,0}\right)\propto{\cal N}\left(\theta_{i,0},m_{\theta_{i,0}},\left(\sigma_{\theta_{i,0}}\right)^2\right)$;\\
  \FOR {$n=1$ to $N_{time}$ (time index)}
           \STATE {nodes} $i\in{\cal M}$ {\bf{in parallel}}
        \STATE ~~~compute the messages from factor vertex $f_i$ to variable vertices $\mu_{{f}_i\to x_{i,n}}(x_{i,n})$, \\~~~$\mu_{{f}_i\to y_{i,n}}(y_{i,n})$, and $\mu_{{f}_i\to \theta_{i,n}}(\theta_{i,n})$ according to \eqref{f_itox_ivmp}-\eqref{f_itotheta_ivmp}.
         \FOR{$l=1$ to $N_{iter}$}
                \STATE ~~\textbf{If} $(i,j)\notin \Omega_n$
                   \STATE  ~~~~~compute all messages from factor vertices to variable vertices $\mu_{f_{ij}\to {\xi}_{i,n}}^{(l)}\left({\xi}_{i,n}\right)$ \\~~~~~according to \eqref{VMPftox}-\eqref{VMPftotheta},  \eqref{VMPftox1}-\eqref{VMPftotheta1} and remove $\hat{b}^{(l-1)}_{ij,n}$;
                   \STATE ~~\textbf{Else}
                    \STATE  ~~~~~compute all messages from factor vertices to variable vertices $\mu_{f_{ij}\to {b}_{ij,n}}^{(l)}\left({\xi}_{i,n}\right)$ and \\~~~~~$\mu_{f_{ij}\to {\xi}_{i,n}}^{(l)}\left({\xi}_{i,n}\right)$ according to \eqref{VMPftox}-\eqref{VMPftotheta1}
                   \STATE ~~update means and variances of beliefs $b^{(l)}\left(\xi_{i,n}\right)$, $b^{(l)}\left(b_{ij,n}\right)$ according to \eqref{beliefb1}-\eqref{vmp_sigmatheta};
                   \STATE ~~broadcast outgoing messages $b^{(l)}\left(\xi_{i,n}\right)$ and $b^{(l)}\left(b_{ij,n}\right)$ to neighboring nodes;
  \ENDFOR;
         \STATE ~~~transmit the belief $b^{(N_{iter})}(x_{i,n})$, $b^{(N_{iter})}(y_{i,n})$ and $b^{(N_{iter})}(\theta_{i,n})$ to the factor vertex $f_i$.
         \STATE ~~~estimate agents' positions and clock offsets using {\bf{MMSE}} estimator;
         \STATE {\bf{end parallel}};
         \ENDFOR;
     \end{algorithmic}
 \end{algorithm}

\subsection{Relationship Between BP-based and VMP-based Algorithms}\label{BPVMP}

The update rules of BP and VMP are presented in the above sections. As mentioned in \cite{yedidia2005constructing}, all message passing methods aim to use the belief $b(\bm{\kappa})$ of a parameter vector $\bm{\kappa}$ to approximate the exact probability distribution function $p(\bm{\kappa})$ which minimize the \emph{Gibbs free energy}. Usually, $b(\bm{\kappa})$ is constrained to be in a class of probability distributions. Using mean-field approximation, $b(\bm{\kappa})$ is fully factorized as $b(\bm{\kappa})=\prod b_i(\kappa_i)$. By minimizing \emph{Gibbs free energy}, the message passing expressions of VMP are obtained \cite{riegler2013merging}. The Bethe method is a region-based approximation $b({\bf{x}})=\prod b_{\bf{a}}(\kappa_{\bf{a}})$ where {\bf{a}} is a subset consisted of different node $i$. Using Bethe approximation and Lagrangian optimization with marginal constraint yields the message expressions of BP \cite{riegler2013merging}. The mean-field approximation constrains all nodes to be independent while Bethe approximation considers the interaction among nodes\cite{opper2001advanced}.

Specifically, considering a likelihood function connecting two variable vertices, the message to one variable vertex depends on the likelihood function and the message from the other variable vertex. In VMP method, the message from the other variable vertex is regarded as the true statistics, therefore the uncertainty of the variable vertex is not taken into account. This can be observed by comparing the variances of the variables' beliefs in BP and VMP, i.e., \eqref{sigmax} and \eqref{vmp_sigmax}-\eqref{vmp_sigmatheta}, respectively. The first term in these expressions are related to the standard deviations of the prior Gaussian distributions $\sigma_{x_{i,0}}$, $\sigma_{y_{i,0}}$ and $\sigma_{\theta_{i,0}}$, which are the same for both BP and VMP. The second terms for VMP only depend on the standard deviation of range measurement $\sigma_d$. In contrast, these terms for BP not only depend on $\sigma_d$, but also relate to the position uncertainties of neighboring nodes. Therefore, we can expect that for small standard deviations of the prior distributions, since the first term dominates the summation of \eqref{sigmax} and \eqref{vmp_sigmax}-\eqref{vmp_sigmatheta}, the difference in terms of localization accuracy by BP and VMP will become negligible. The performance of different algorithms will be evaluated in Section \ref{simulation}.

\subsection{Message Passing Schedule}\label{scheme}
It can be observed from Fig. \ref{Fig33} and Fig. \ref{Fig4} that the FGs contain cycles. Hence, both the proposed BP-based and VMP-based algorithm are iterative and different message passing schedules have to be considered. In the proposed \textbf{Algorithm} \ref{algorithm1} and \textbf{Algorithm} \ref{algorithm2}, agents update messages related to their own variables and transmit them to neighbors via wireless communications. Since the wireless transmission is performed at each iteration, communication overhead of this message passing schedule is proportional to the total number of iterations $N_{iter}$. We propose another message passing schedule, in which agents perform more than one iterations to update the outgoing messages before they transmit them to neighbors. This message passing schedule consists of two iteration loops, i.e., internal iteration which is performed locally by an agent, and external iteration which exchanges information between neighbors. Obviously, by designing of the number of internal iterations $N_{int}$ and that of the external iterations $N_{ext}$ properly, it is able to reduce the communication overhead given the total number of iterations $N_{iter}=N_{int} N_{ext}$.

Specifically, for a certain time $n$ and a pair of nodes $(i,j)\in{\Xi}$, at the $(l-1)$-{th} external iteration and the $p$-{th} internal iteration, we can calculate the messages $\mu_{f_{ij}\to {x}_{i,n}}^{(l-1)(p)}\left({x}_{i,n}\right)$, $\mu_{f_{ij}\to {y}_{i,n}}^{(l-1)(p)}\left({y}_{i,n}\right)$ and $\mu_{f_{ij}\to {\theta}_{i,n}}^{(l-1)(p)}\left({\theta}_{i,n}\right)$ as\footnote{We only illustrate the messages of LOS measurement for brevity. The results in NLOS conditions are straightforward.}
\begin{align}\label{dispatchf_to_xi}
\mu_{f_{ij}\to {x}_{i,n}}^{(l-1)(p)}\left({x}_{i,n}\right)&=\int\int{f_{ij}}\cdot\mu^{(l-1)(p-1)}_{y_{i,n}\to f_{ij}}\left(y_{i,n}\right)\mu^{(l-1)(p-1)}_{\theta_{i,n}\to f_{ij}}\left(\theta_{i,n}\right)\mathrm{d} y_{i,n} \mathrm{d} \theta_{i,n},\\
\mu_{f_{ij}\to {y}_{i,n}}^{(l-1)(p)}\left({y}_{i,n}\right)&=\int\int{f_{ij}}\cdot\mu^{(l-1)(p-1)}_{x_{i,n}\to f_{ij}}\left(x_{i,n}\right)\mu^{(l-1)(p-1)}_{\theta_{i,n}\to f_{ij}}\left(\theta_i\right)\mathrm{d} x_{i,n} \mathrm{d} \theta_{i,n},
\\\label{dispatchf_to_thetai}
\mu_{f_{ij}\to {\theta}_{i,n}}^{(l-1)(p)}\left({\theta}_{i,n}\right)&=\int\int{f_{ij}}\cdot\mu^{(l-1)(p-1)}_{x_{i,n}\to f_{ij}}\left(x_{i,n}\right)\mu^{(l-1)(p-1)}_{y_{i,n}\to f_{ij}}\left(y_{i,n}\right)\mathrm{d} x_{i,n} \mathrm{d} y_{i,n}.
\end{align}

Then the updated messages at the $p$-{th} internal iteration are calculated by
\begin{align}\label{dispatchx_to_f}
    \mu^{(l-1)(p)}_{{\xi}_{i,n}\to f_{ij}}\left({\xi}_{i,n}\right)=\prod_{j^{'}\in {\cal S}_{i,n}\backslash j}\mu^{(l-1)(p)}_{f_{ij^{'}}\to \xi_{i,n}}\left({\xi}_{i,n}\right).
\end{align}

The internal iteration repeats by performing \eqref{dispatchf_to_xi}-\eqref{dispatchx_to_f} iteratively, until the number of internal iterations reaches the maximum value $N_{int}$. Then, at the $l$-{th} external iteration, the outgoing messages of agent $i$ to be transmitted to agent $j$ are obtained by
\begin{align}\label{schemex}
\mu_{{\xi}_{i,n} \to f_{ij}}^{(l)}\left({\xi}_{i,n}\right)=\mu_{{\xi}_{i,n} \to f_{ij}}^{(l-1)(N_{int})}\left({\xi}_{i,n}\right).
\end{align}
At the same time, agent $i$ receives messages from its neighbors and begins a new round of internal iterations. When the number of external iterations reaches the maximum value $N_{ext}$, the message passing stops and agents are able to calculate the beliefs of their own variables separately. The message passing schedule of the proposed VMP algorithm is similar to that of the BP algorithm. The message passing schedule for \emph{standard} BP is characterized in \textbf{Algorithm 3} and its performance with different parameters are going to be evaluated by simulations.

\begin{table}[!t]
\label{table11}
\normalsize
\hrule
\vskip 0.1cm
\center
{{\bf\normalsize{Algorithm 3. Message Passing Schedule}}} \\
\vskip 0.1cm
\hrule
\vskip 0.1cm
\begin{enumerate}
\item[] At time $n$
  \item[1:] {\bf{for}} $l=1$ to $N_{ext}$ (External iterations)
  \item[2:] ~~~~{nodes} $i\in{\cal M}$ {\bf{in parallel}}
  \item[3:] ~~~~~~compute all messages from factor vertex to variable vertex $\mu_{f_{ij}\to {x}_{i,n}}^{(l)}\left({x}_{i,n}\right)$, \\~~~~~~$\mu_{f_{ij}\to {y}_{i,n}}^{(l)}\left({y}_{i,n}\right)$ and $\mu_{f_{ij}\to {\theta}_{i,n}}^{(l)}\left({\theta}_{i,n}\right)$,
  \item[4:] ~~~~~~{Entering the {\bf internal }iteration loop}:
  \item[5:] ~~~~~~Initialize $\mu_{{x}_{i,n} \to f_{ij}}^{(l)(0)}\left({x}_{i,n}\right)=\mu_{{x}_{i,n} \to f_{ij}}^{(l)}\left({x}_{i,n}\right)$, \\~~~~~~~~~~~~~~~~~$\mu_{{y}_{i,n} \to f_{ij}}^{(l)(0)}\left({y}_{i,n}\right)=\mu_{ {y}_{i,n} \to f_{ij}}^{(l)}\left({y}_{i,n}\right)$, \\~~~~~~~~~~~~~~~~~$\mu_{{\theta}_{i,n} \to f_{ij}}^{(l)(0)}\left({\theta}_{i,n}\right)=\mu_{ {\theta}_{i,n} \to f_{ij}}^{(l)}\left({\theta}_{i,n}\right)$
  \item[6:] ~~~~~~~~~{\bf{for}} $p=1$ to $N_{int}$ (Internal iterations) 
  \item[7:] ~~~~~~~~~~~calculate the incoming messages:
  \item[] ~~~~~~~~~~~~~~$\mu_{f_{ij}\to {x}_{i,n}}^{(l)(p)}\left({x}_{i,n}\right)$, $\mu_{f_{ij}\to {y}_{i,n}}^{(l)(p)}\left({y}_{i,n}\right)$, $\mu_{f_{ij}\to {\theta}_{i,n}}^{(l)(p)}\left({\theta}_{i,n}\right)$ according to \eqref{dispatchf_to_xi}-\eqref{dispatchf_to_thetai};
  \item[8:] ~~~~~~~~~~~and then update messages:
  \item[] ~~~~~~~~~~~~~~$\mu_{{x}_{i,n}\to f_{ij}}^{(l)(p)}\left({x}_{i,n}\right)$, $\mu_{{y}_{i,n}\to f_{ij}}^{(l)(p)}\left({y}_{i,n}\right)$, $\mu_{{\theta}_{i,n}\to f_{ij}}^{(l)(p)}\left({\theta}_{i,n}\right)$ according to \eqref{dispatchx_to_f};
  \item[10:] ~~~~~~~~{\bf end for}
  \item[11:] ~~~~~~update the outgoing messages $\mu_{ {x}_{i,n} \to f_{ij}}^{(l+1)}\left({x}_{i,n}\right)$, $\mu_{{y}_{i,n}\to f_{ij}}^{(l+1)}\left({y}_{i,n}\right)$, $\mu_{{\theta}_{i,n} \to f_{ij}}^{(l+1)}\left({\theta}_{i,n}\right)$ according \\~~~~~~to \eqref{schemex};
  \item[12:] ~~~~~~transmit the outgoing messages to neighboring nodes;
  \item[13:] ~~~~~~calculate the beliefs and perform MMSE estimation of variables according to \\~~~~~~\textbf{Algorithm}. 1.
  \item[14:] ~~~~~{\bf{end parallel}}
  \item[15:] {\bf end for}
\end{enumerate}
\vskip 0.1cm
\hrule
\end{table}

\section{Simulation Results and Discussions}\label{simulation}
\begin{figure}
\centering
\includegraphics[width=.65\textwidth]{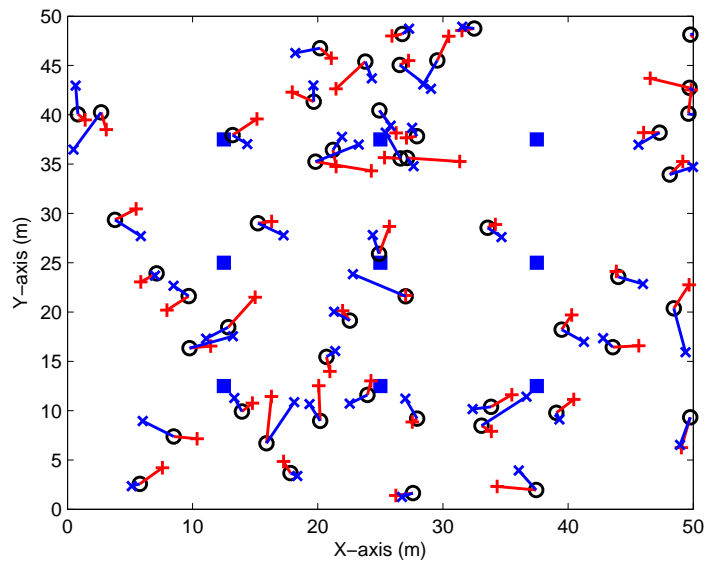}
\caption{\quad The single trail localization results of the proposed algorithms. The red line shows the location error of the proposed \emph{standard} BP and the blue line denote the location error of the proposed VMP.}\label{Fig6}
\centering
\end{figure}

We evaluate the performance of the proposed BP and VMP algorithms for distributed cooperative joint localization and synchronization. Consider a $50\times50m^2$ plane with 9 static anchor nodes and 50 mobile agent nodes, as illustrated in Fig. \ref{Fig6}. Anchors denoted by `$\blacksquare$' are synchronized and have perfect knowledge of their positions. Agents denoted by `$\bigcirc$' are uniformly distributed on the plane. The maximum communication range is set to $20m$, i.e., $(i,j)\in{\Xi}$ and $(j,i)\in{\Xi}$ if and only if $\|\bm{x}_i-\bm{x}_j\|\leq 20m$. It is assumed that clock offsets are uniformly distributed and the maximum equivalent range offset due to the clock offset is $c\cdot\theta=50m$. The prior distributions of agents' positions are assumed to be Gaussian with variances $\sigma_{x_{i,0}}^2=\sigma_{y_{i,0}}^2=100 m^2$. The state transition noise is assumed to be identical at different time slots, i.e., $\sigma_{ux,n}=\sigma_{uy,n}=1m$, $\sigma_{u\theta,n}=10ns$, $\forall n$. The velocity of agent node on $x$-axis and $y$-axis are uniformly generated from $[0,3]~m/s$ and time interval $\Delta_t$ is set to $1s$. The range measurement noise are zero mean Gaussian distributed with variance $\sigma_d^2=c^2 \sigma_t^2= 1m^2$, and the maximum number of iterations is set to $N_{iter}=20$, unless otherwise specified. In the following, we will first evaluate the performance of the proposed algorithms in LOS environment. Then, mixed LOS/NLOS conditions will also be studied.

\subsection{Joint Localization and Synchronization Accuracy}
\begin{figure}
\centering
\includegraphics[width=.65\textwidth]{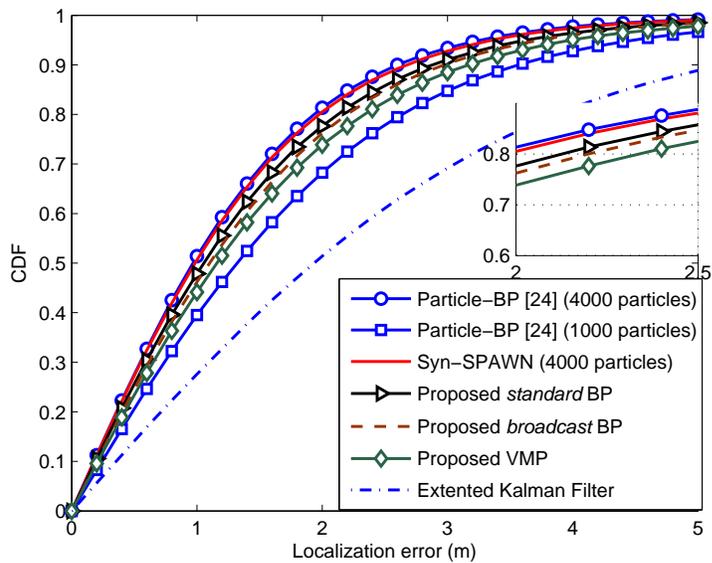}
\caption{CDFs of localization error of various algorithms.}\label{Fig7}
\centering
\end{figure}

A single trail localization results of the proposed \emph{standard} BP and VMP algorithms are illustrated in Fig.~\ref{Fig6}, denoted by `$+$' and `$\times$', respectively. It is seen that both the localization results of the two algorithms are close to the true positions of agents. The cumulative distribution function (CDF) of localization errors of the proposed algorithms at time $n=10$ are compared with that of that of three state-of-the-art methods, namely, `Syn-SPAWN' and `Particle-BP' \cite{meyer2013distributed} and extended Kalman filter in Fig.~\ref{Fig7}\footnote{In the following simulation results, we choose the time $n=10$ unless otherwise specified.}. Syn-SPAWN' denotes the combined method which performs network synchronization \cite{leng2011distributed} followed by cooperative localization using SPAWN \cite{wymeersch2009cooperative}. Syn-SPAWN and particle-BP are implemented by using 4000 and 1000 particles to represent the messages on FG. The extended Kalman Filter method treats the uncertainties of nodes' positions and clock offsets as measurement noise and thus suffers performance degradation. The proposed BP and VMP joint localization and synchronization algorithms perform very close to the existing particle-based methods with much less communication overhead thanks to the parametric message representation. They perform even better than the particle-based methods when the number of particles is insufficient. Communication overhead of the proposed BP algorithm can be further reduced by applying the proposed \emph{broadcast} BP scheme, which approximates the outgoing messages of a variable vertex by its beliefs. We can observe from Fig.~\ref{Fig7} that the approximation leads to negligible performance loss.

%

%

\begin{figure}
\centering
\includegraphics[width=.65\textwidth]{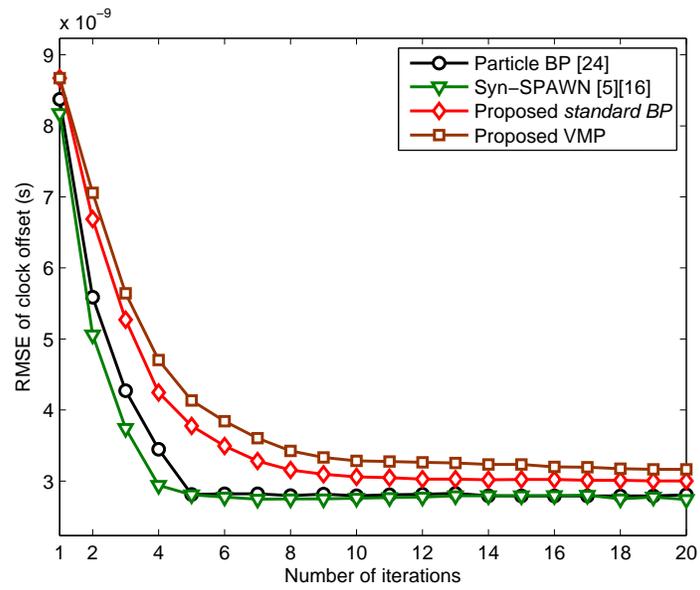}
\caption{RMSEs of clock offset versus iterations.}\label{Fig8}
\end{figure}

\begin{figure}
\centering
\includegraphics[width=.65\textwidth]{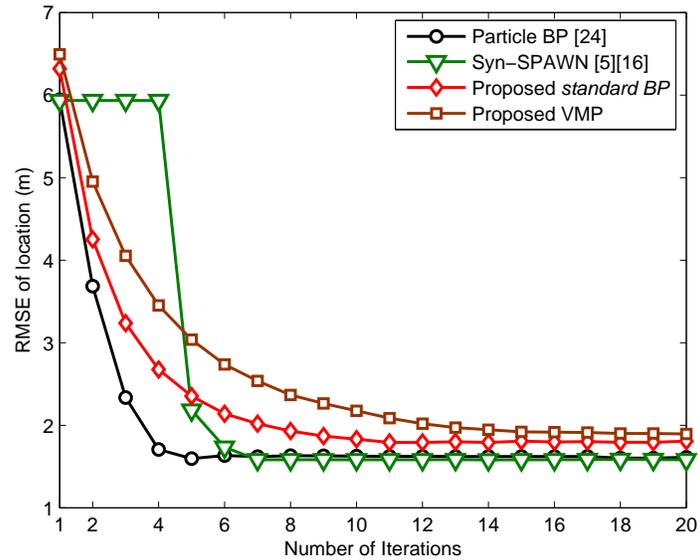}
\caption{RMSEs of location versus iterations.}\label{Fig9}
\end{figure}

The RMSE of the estimated clock offsets versus the number of iterations of the proposed algorithms, particle-BP and Syn-SPAWN are shown in Fig.~\ref{Fig8}. All the four algorithms converge in a few iterations. However, the BP synchronization method in \cite{leng2011distributed} requires a round-trip timing mechanism. On the contrary, the proposed algorithms utilize the one way measurement and perform localization and synchronization simultaneously. The RMSEs of the location estimations of the proposed algorithms, Syn-SPAWN and particle-BP are plotted in Fig.~\ref{Fig9}. Since Syn-SPAWN performs synchronization before cooperative localization, RMSE of location holds for several iterations. The proposed BP and VMP algorithms improve the position accuracy from the first iteration. After the convergence, the two algorithms can almost attach Syn-SPAWN.

\begin{figure}
\centering
\includegraphics[width=.65\textwidth]{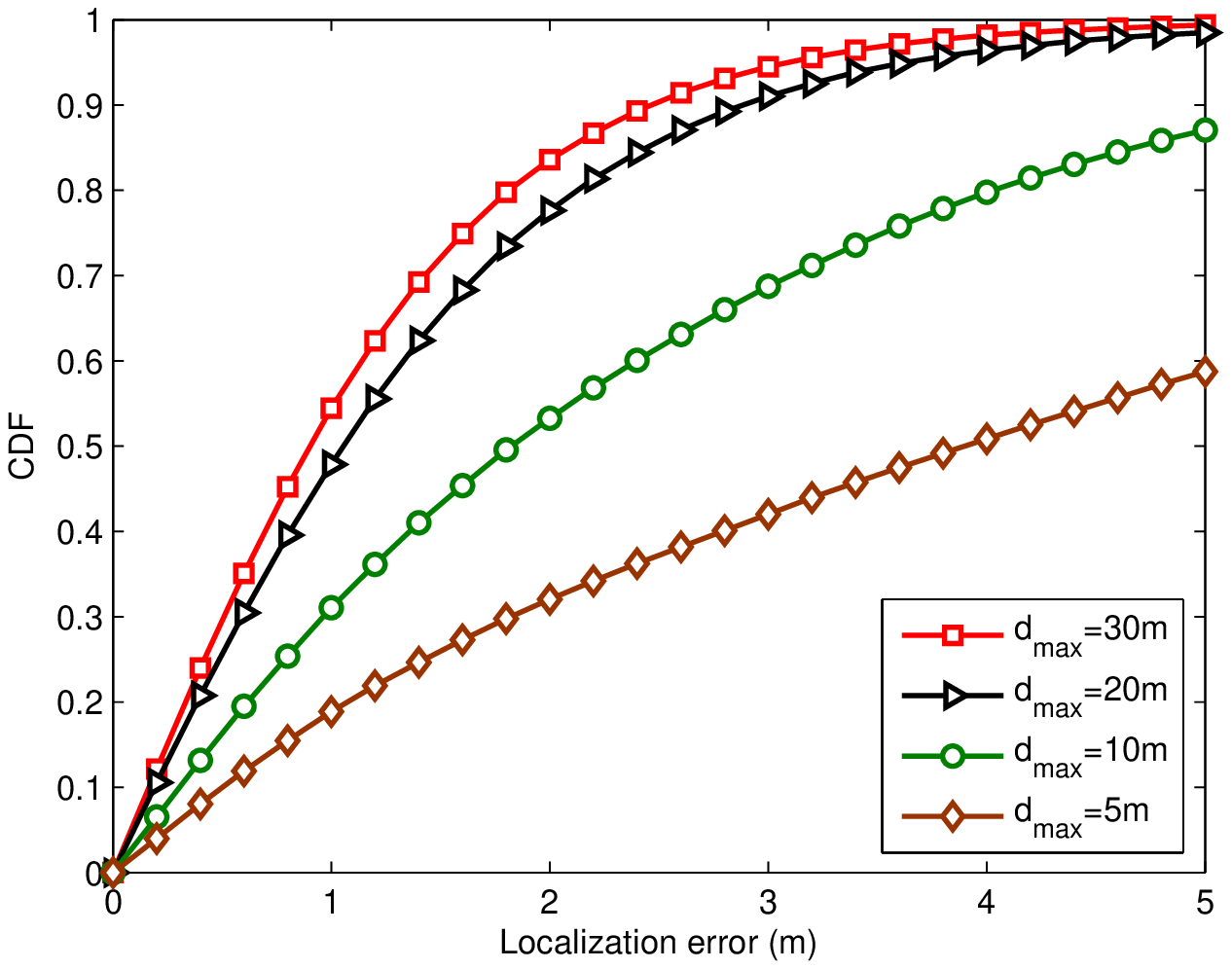}
\caption{CDFs of the proposed algorithms with different communication ranges}\label{Fig12}
\centering
\end{figure}

The averaged number of cooperative nodes for a given network topology depends on the maximum communication range. Fig.~\ref{Fig12} shows the accuracy of the proposed \emph{standard} BP location estimation with communication ranges $d_{max}=5m$, $d_{max}=10m$, $d_{max}=20m$ and $d_{max}=30m$. We can observe that localization performance improves as the increase of communication range. However, the improvement becomes negligible when the maximum communication range is large enough. On the other hand, in order to obtain the same accuracy of range measurement, signal power has to be increased exponentially as the increase of communication range. Therefore we can trade off between localization accuracy and power cost in wireless transmission.

\begin{figure}[]
\centering
{\includegraphics[width=.65\textwidth]{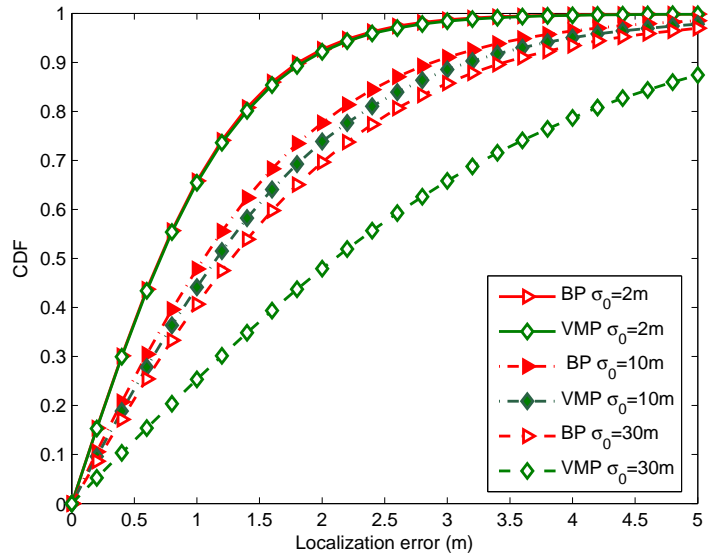}
\label{Fig13b}}
\caption{Impact of prior uncertainty on the proposed BP and VMP}\label{Fig13}
\end{figure}

\begin{figure}[]
\centering
{\includegraphics[width=.65\textwidth]{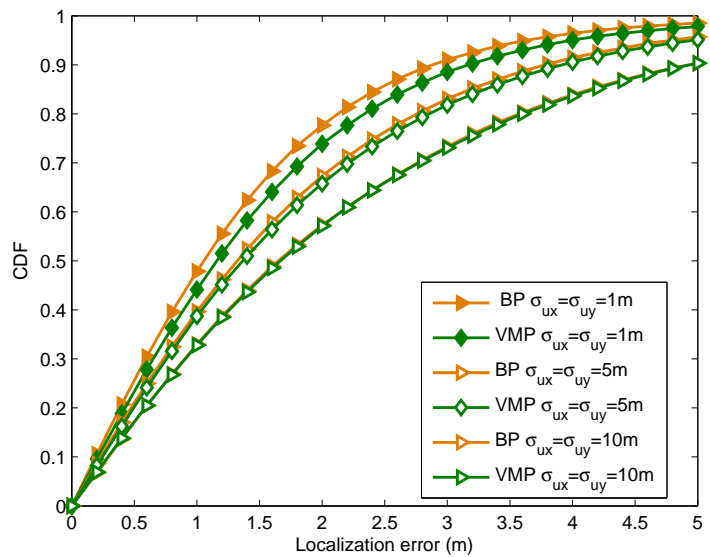}
\label{Fig13b}}
\caption{Impact of state transition noise on the proposed BP and VMP}\label{noise}
\end{figure}

It is seen from the above results that the proposed BP and VMP algorithms perform very close to each other given the standard deviations of the prior distribution of agent's position $\sigma_{x_{i,0}}=\sigma_{y_{i,0}}=10m$. We can also observe from \eqref{VMPftox}-\eqref{VMPftotheta1} that, in message updating of VMP, uncertainties of neighboring nodes are neglected. To further clarify this difference, the CDFs of localization error for different prior uncertainties of nodes' positions are shown in Fig.~\ref{Fig13}. It is seen that for small prior uncertainty, e.g., $\sigma_{x_{i,0}}=\sigma_{y_{i,0}}=2m$, localization performance of the proposed BP and VMP are very close. As the prior uncertainty increases, the performance gap between the proposed BP and VMP becomes larger. It can be seen that for $\sigma_{x_{i,0}}=\sigma_{y_{i,0}}=30m$, BP significantly outperforms VMP. In addition, comparing \eqref{f_itox_i}-\eqref{f_itotheta_i} and \eqref{f_itox_ivmp}-\eqref{f_itotheta_ivmp}, it can be found that the uncertainties of nodes' beliefs are also neglected in VMP. Since the uncertainty of node's belief varies, we plot the CDFs of localization error for different standard deviations of transition noise in Fig.~ \ref{noise}. When the variance of transition noise is small, VMP which omits the uncertainty of belief leads to performance degradation. However, when the variance of transition noise is large enough, the performance loss of VMP is negligible. The simulation results corroborate the discussion about the relationship between the proposed BP and VMP for localization in the previous section.



\subsection{Message Passing Schedule}\label{compexity}

\begin{figure}[]
\centering
\includegraphics[width=.65\textwidth]{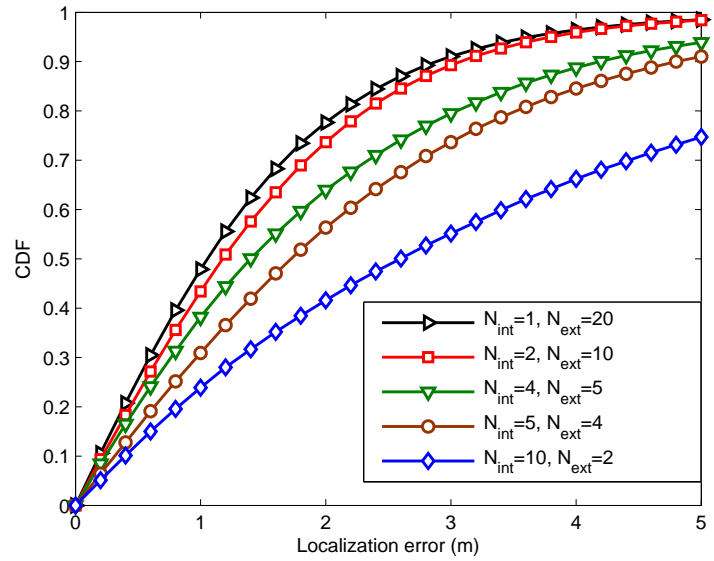}
\caption{Different message passing schedules with the same complexity }
\label{schedulCDF1}
\end{figure}

We evaluate the performance of different message passing schedules for the proposed \emph{standard} BP. Two cases are going to be studied. In the first case, the total number of iterations is set to $N_{iter}=20$, and the number of internal iterations $N_{int}$ and that of external iterations $N_{ext}$ vary accordingly. Five pairs of configurations are considered: $(1) N_{int}=1, N_{ext}=20$, $(2) N_{int}=2, N_{ext}=10$, $(3) N_{int}=4, N_{ext}=5$, $(4) N_{int}=5, N_{ext}=4$ and $(5) N_{int}=10, N_{ext}=2$. Since the total number of message update depends on $N_{iter}=N_{int}N_{ext}$, computational complexities of all the above configurations are identical. Simulation results are shown in Fig.~\ref{schedulCDF1}. We can observe that, given $N_{iter}=20$, localization performance degrades as $N_{ext}$ decreases. This phenomenon reveals that message passing between neighboring nodes are more informative than that between its own variables. Nevertheless, since the number of wireless transmission between nodes depends on $N_{ext}$, a small value of $N_{ext}$ results in lower communication overhead, which is important especially in dense network.

\begin{figure}[]
\centering
\includegraphics[width=.65\textwidth]{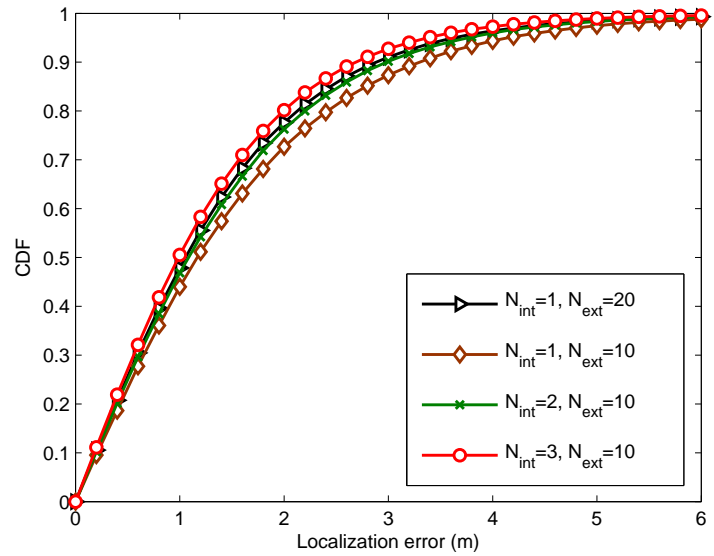}
\caption{The impact of internal iteration on localization accuracy }
\label{schedulCDF2}
\end{figure}

In the second case, we study the impact of $N_{int}$ on the localization performance. Four pairs of configurations are considered: $(1) N_{int}=1, N_{ext}=20$, $(2) N_{int}=1, N_{ext}=10$, $(3) N_{int}=2, N_{ext}=10$ and $(4) N_{int}=3, N_{ext}=10$. The results are illustrated in Fig.~\ref{schedulCDF2}. We can observe that the first configuration outperforms the second and the third one, even though the first and the third configurations have the same $N_{iter}$. This interesting phenomenon again demonstrates the importance of external iteration compared to the internal iteration. Nevertheless, when we keep increasing the number of internal iterations, localization accuracy improves. By comparing the first and the last configurations, we can see that the last one outperforms the first one at the cost of more computational complexities. However, the last configuration has a smaller number of $N_{ext}$, which means less communication overhead. Therefore, in practical applications, localization accuracy, communication overhead and computational complexity can be compromised by designing the message passing schedule on the FG.

\begin{figure}[]
\centering
{\includegraphics[width=.65\textwidth]{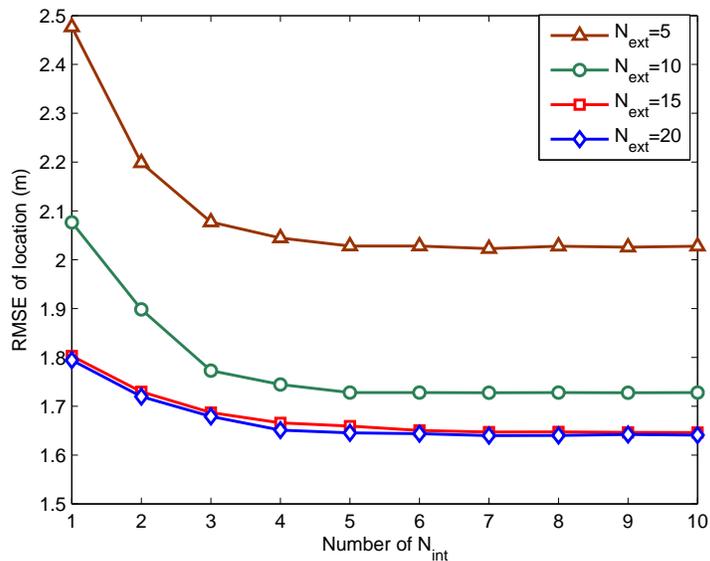}
\label{Fig13b}}
\caption{RMSE of location versus number of internal iterations }\label{schedule}
\end{figure}

The RMSE of localization versus $N_{int}$ is plotted in Fig.~\ref{schedule}. It is seen that, for different values of $N_{ext}$, location error converges after $N_{int}=5$. Moreover, we can observe that increasing the value of $N_{ext}$ can improve the location estimation accuracy. However, the performance improvement becomes negligible when $N_{ext}$ is greater than $15$.


\subsection{Performance of the Proposed Algorithms in Mixed LOS/NLOS Environment}

\begin{figure}
\centering
\includegraphics[width=.65\textwidth]{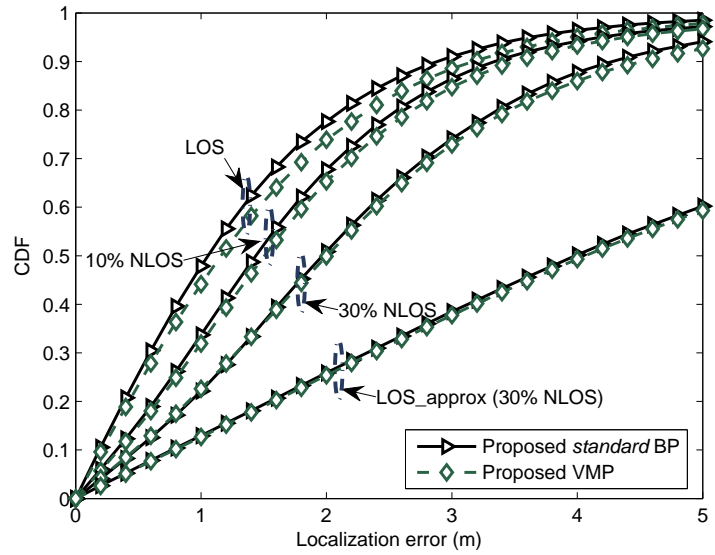}
\caption{CDFs of localization error in mixed LOS/NLOS environments.}\label{Nloslos}
\centering
\end{figure}

\rd{The CDF of localization error of the proposed algorithms in a mixed LOS/NLOS environment is illustrated in Fig.~\ref{Nloslos} with different percentage of NLOS connections. The rate parameter for NLOS bias is $\lambda=0.38 m^{-1}$. For comparison purpose, the result by treating all measurements as LOS, denoted as ``LOS-approx.'', is also plotted. We can observe that NLOS measurements degrade the localization performance significantly if the bias is not taken into account in the algorithms. Using the proposed algorithms, the impact of NLOS can be notablely relieved. Moreover, the performance gap between the proposed BP and VMP becomes smaller when the percentage of NLOS measurements increases. This is due to the fact that the moment matching applied in BP to approximate messages related to NLOS measurement leads to performance loss, which becomes noticeable with more NLOS links.}

\begin{figure}
\centering
\includegraphics[width=.7\textwidth]{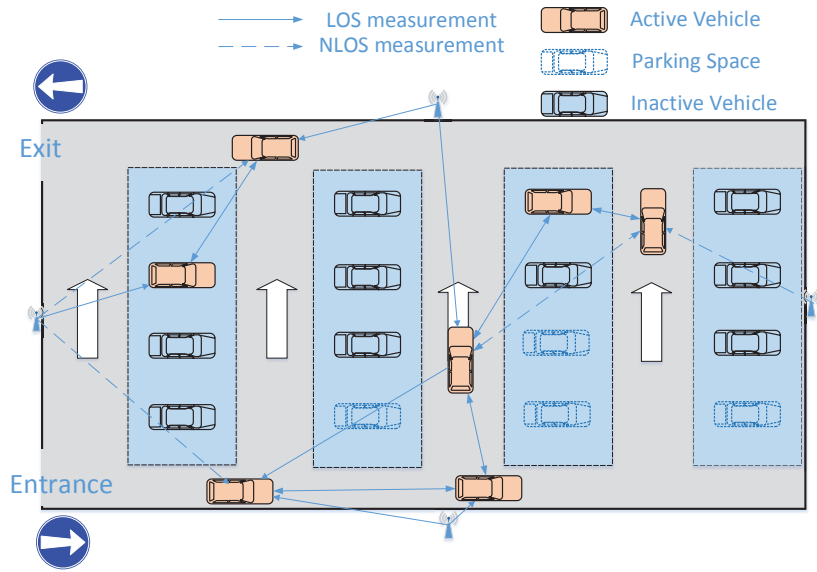}
\caption{Parking floor environment.}\label{park}
\centering
\end{figure}

\begin{figure}
\centering
\includegraphics[width=.7\textwidth]{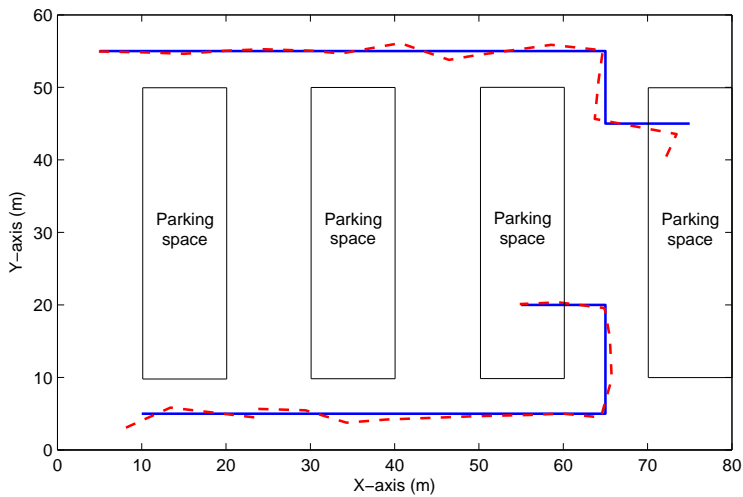}
\caption{The true trajectories (blue solid lines) and estimated trajectories based on the proposed \emph{standard} BP (red dashed lines).}\label{path}
\centering
\end{figure}

\begin{figure}
\centering
\includegraphics[width=.65\textwidth]{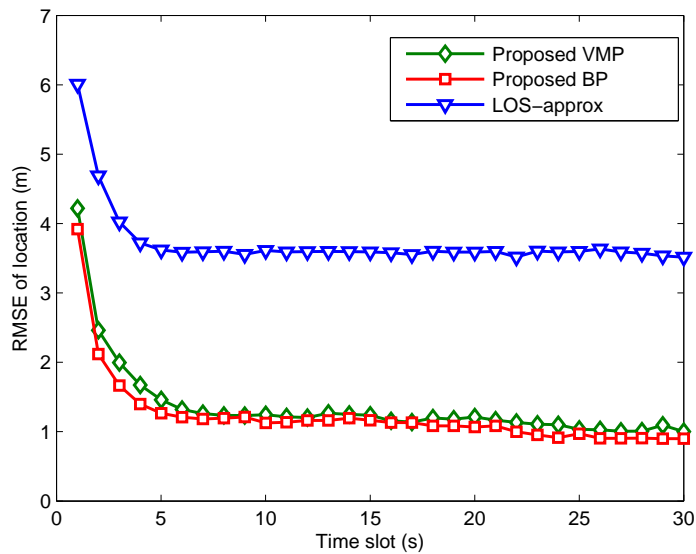}
\caption{RMSEs of location versus time in the parking floor scenario.}\label{rmse}
\centering
\end{figure}

\rd{We further consider a practical scenario of vehicle localization in a $80\times 60 m^2$ parking floor, which is shown in Fig.~\ref{park}. In this situation, GPS signal is weak or even unavailable. Four anchor nodes are located against the walls. There are inactive vehicles parked in the parking lots, which do not participate in cooperative localization. Several active vehicles that can cooperate with each other are either looking for parking space or leaving the parking floor. The speed of vehicle is set to $v<10 m/s$. The communication range of each node is $50m$. We assume that if the direct path between two active vehicles or between vehicles and anchor nodes within the communication range are been obstructed, NLOS measurement is obtained for this link, denoted by dashed arrows. Otherwise, the measurement is LOS, denoted by solid arrows. The other parameters are the same as that of the previous simulations. In Fig.~\ref{path}, the true and estimated trajectories of two chosen vehicles are plotted. It is seen that the estimated trajectories by the proposed algorithms are close to the true trajectories of vehicles, which validates the proposed algorithms. The RMSEs of location estimated by the proposed algorithms are illustrated in Fig.~\ref{rmse}. It is seen that all the algorithms converge in 5-10 seconds. Due to the rich NLOS components in this scenario, the proposed BP and VMP algorithms significantly outperform the one that is not aware of NLOS, which further demonstrates the superior performance of the proposed algorithms in practical scenarios. }

\subsection{Computational Complexity and Communication Overhead Analysis}\label{compexity}

We analyze the computational complexity and communication overhead of the proposed algorithms and that of the state-of-the-art methods in LOS environment\footnote{Computational complexity and communication overhead increase for both the proposed algorithms and the state-of-the-art methods in NLOS environment. Therefore, for brevity, we only compare the results in LOS condition in this paper.}. Communication overhead is evaluated by the number of parameters transmitted to neighbors. As the algorithms are distributed, the computation is calculated at each agent node. Therefore we consider the computation complexity of one agent node.

For the particle-based Syn-SPAWN using $R$ particles, the complexity consists of the operations in both localization and synchronization, which scales as ${\cal O}(R)+{\cal O}({{R^2}N(i)})+{\cal O}(N(i))$ and the number of parameters broadcast per iteration is ${\cal O}(R)+2{\cal O}(1)$\cite{lien2012comparison}. The particle-based joint estimation method achieve localization and synchronization simultaneously, so the complexity only depends on the number of particles $R$ \cite{meyer2013distributed}. The complexity is ${\cal O}(R)+{\cal O}({{R}N(i)})$ and communication overhead is ${\cal O}(R)$. For the proposed \emph{standard} BP, an agent node calculates different messages to neighbors, the number of operations scales as ${\cal O}(N^2(i))$. For the proposed \emph{broadcast} BP and VMP algorithms, due to the broadcast feature, the operation complexity reduces to the scale of ${\cal O}(N(i))$. Because VMP algorithm approximate the posterior distribution with independent beliefs, only means are sent to neighbors. Therefore, the communication overhead of the proposed VMP is $3{\cal O}(1)$, while that of the proposed \emph{standard} BP is $6{\cal O}(N(i))$. In the proposed \emph{broadcast} BP method each node broadcast its belief, so the communication overhead is down to $1/N(i)$ of the \emph{standard} BP. The comparison of complexity and communication overhead for different algorithms is summarized in Table \ref{table2}, which shows that the proposed VMP algorithm has the lowest computational complexity and communication overhead.

\begin{table*}[!t]
\caption{Comparison of computational complexity and communication overhead of one agent node}
\centering
\label{table2}
\begin{tabular}{|c|c|c|c|c|}
\hline
 Algorithm &  Complexity & Transmit Parameters \\
\hline
 Syn-SPAWN \cite{leng2011distributed}\cite{wymeersch2009cooperative}& ${\cal O}(R)+{\cal O}({{R^2}N(i)})+{\cal O}(N(i))$ & ${\cal O}(R)+2\cdot{\cal O}(1)$ \\
\hline
 Particle-BP \cite{meyer2013distributed} & ${\cal O}(R)+{\cal O}({{R}N(i)})$ & ${\cal O}(R)$  \\
\hline
\emph{Standard} BP algorithm & ${\cal O}(N^2 (i))$ & $6\cdot{\cal O}(N(i))$  \\
\hline
\emph{Broadcast} BP algorithm & ${\cal O}(N(i))$ & $6\cdot{\cal O}(1)$  \\
\hline
Proposed VMP algorithm & ${\cal O}(N(i))$ & $3\cdot{\cal O}(1)$  \\
\hline
\end{tabular}
\end{table*}


\section{Conclusions}
In this paper, a unified factor graph framework was proposed to solve the distributed joint cooperative localization and synchronization problem in dynamic wireless networks. We investigated BP and VMP algorithms based on TOA measurements in line-of-sight (LOS) and non-line-of-sight (NLOS) environments. The messages of BP and VMP were intractable to be expressed in closed forms due to the nonlinear terms in the observation function. For this reason, Taylor expansion was used to linearize the specific nonlinear term in the expressions of messages. Accordingly, all the messages on FG were derived in Gaussian forms and only means and variances were required to be updated and transmitted. Based on the observation that there are two iteration loops, namely, internal iteration and external iteration, a message passing schedule was proposed to trade off between the number of information exchange with neighboring nodes and the estimation precision. Simulation results showed that the proposed joint estimation algorithms performed very close to the particle-based algorithm with much lower communication overhead and complexity. The proposed BP and VMP algorithms performed close to each other when the position uncertainties of neighboring nodes are negligible, in which case VMP can be more attractive in practice due to the enabling of broadcast transmission.

\section*{Acknowledgement}
The authors would like to thank the editor and the anonymous reviewers for their valuable comments which significantly improved the quality of this paper.


\bibliographystyle{IEEEtran}
\bibliography{IEEEabrv,bib}

\begin{thebibliography}{10}
\providecommand{\url}[1]{#1}
\csname url@samestyle\endcsname
\providecommand{\newblock}{\relax}
\providecommand{\bibinfo}[2]{#2}
\providecommand{\BIBentrySTDinterwordspacing}{\spaceskip=0pt\relax}
\providecommand{\BIBentryALTinterwordstretchfactor}{4}
\providecommand{\BIBentryALTinterwordspacing}{\spaceskip=\fontdimen2\font plus
\BIBentryALTinterwordstretchfactor\fontdimen3\font minus
  \fontdimen4\font\relax}
\providecommand{\BIBforeignlanguage}[2]{{%
\expandafter\ifx\csname l@#1\endcsname\relax
\typeout{** WARNING: IEEEtran.bst: No hyphenation pattern has been}%
\typeout{** loaded for the language `#1'. Using the pattern for}%
\typeout{** the default language instead.}%
\else
\language=\csname l@#1\endcsname
\fi
#2}}
\providecommand{\BIBdecl}{\relax}
\BIBdecl

\bibitem{gezici2005localization}
S.~Gezici, Z.~Tian, G.~B. Giannakis, H.~Kobayashi, A.~F. Molisch, H.~V. Poor,
  and Z.~Sahinoglu, ``Localization via ultra-wideband radios: a look at
  positioning aspects for future sensor networks,'' \emph{{IEEE} Signal
  Process. Mag.}, vol.~22, no.~4, pp. 70--84, 2005.

\bibitem{misra2006global}
P.~Misra and P.~Enge, \emph{Global Positioning System: Signals, Measurements
  and Performance Second Edition}.\hskip 1em plus 0.5em minus 0.4em\relax
  Massachusetts: Ganga-Jamuna Press, 2006.

\bibitem{patwari2005locating}
N.~Patwari, J.~N. Ash, S.~Kyperountas, A.~O. Hero, R.~L. Moses, and N.~S.
  Correal, ``Locating the nodes: cooperative localization in wireless sensor
  networks,'' \emph{{IEEE} Signal Process. Mag.}, vol.~22, no.~4, pp. 54--69,
  2005.

\bibitem{tseng2009hybrid}
P.-H. Tseng and K.-T. Feng, ``Hybrid network/satellite-based location
  estimation and tracking systems for wireless networks,'' \emph{{IEEE} Trans.
  Veh. Technol.}, vol.~58, no.~9, pp. 5174--5189, 2009.

\bibitem{wymeersch2009cooperative}
H.~Wymeersch, J.~Lien, and M.~Z. Win, ``Cooperative localization in wireless
  networks,'' \emph{Proc. {IEEE}}, vol.~97, no.~2, pp. 427--450, 2009.

\bibitem{nguyen2015least}
T.~V. Nguyen, Y.~Jeong, H.~Shin, and M.~Z. Win, ``Least square cooperative
  localization,'' \emph{{IEEE} Trans. Veh. Technol.}, vol.~64, no.~4, pp.
  1318--1330, 2015.

\bibitem{6476037}
V.~Ekambaram, K.~Ramachandran, and R.~Sengupta, ``Collaborative high accuracy
  localization in mobile multipath environments,'' \emph{{IEEE} Trans. Veh.
  Technol.}, vol.~62, no.~3, pp. 1--8, 2013.

\bibitem{vakulya2011fast}
G.~Vakulya and G.~Simon, ``Fast adaptive acoustic localization for sensor
  networks,'' \emph{{IEEE} Trans. Instrum. Meas.}, vol.~60, no.~5, pp.
  1820--1829, 2011.

\bibitem{chan2006time}
Y.-T. Chan, W.-Y. Tsui, H.-C. So, and P.-c. Ching, ``Time-of-arrival based
  localization under nlos conditions,'' \emph{{IEEE} Trans. Veh. Technol.},
  vol.~55, no.~1, pp. 17--24, 2006.

\bibitem{gustafsson2003positioning}
F.~Gustafsson and F.~Gunnarsson, ``Positioning using time-difference of arrival
  measurements,'' in \emph{Proc. IEEE Int. Conf. on Acous., Speech, and Sig.
  Process.}, vol.~6, 2003, pp. VI--553.

\bibitem{vossiek2003wireless}
M.~Vossiek, L.~Wiebking, P.~Gulden, J.~Wieghardt, C.~Hoffmann, and P.~Heide,
  ``Wireless local positioning,'' \emph{{IEEE} Microw. Mag.}, vol.~4, no.~4,
  pp. 77--86, 2003.

\bibitem{del2003link}
J.~del Prado~Pavon and S.~Choi, ``Link adaptation strategy for ieee 802.11 wlan
  via received signal strength measurement,'' in \emph{Proc. IEEE Int. Conf. on
  Commun.}, vol.~2, 2003, pp. 1108--1113.

\bibitem{ganeriwal2003timing}
S.~Ganeriwal, R.~Kumar, and M.~B. Srivastava, ``Timing-sync protocol for sensor
  networks,'' in \emph{Proc. the 1st ACM Int. Conf. on Embed. Net. Sens.
  Syst.}, 2003, pp. 138--149.

\bibitem{maroti2004flooding}
M.~Mar{\'o}ti, B.~Kusy, G.~Simon, and {\'A}.~L{\'e}deczi, ``The flooding time
  synchronization protocol,'' in \emph{Proc. of the 2nd ACM Int. Conf. on
  Embed. Net. Sens. Syst.}, 2004, pp. 39--49.

\bibitem{schenato2007distributed}
L.~Schenato and G.~Gamba, ``A distributed consensus protocol for clock
  synchronization in wireless sensor network,'' in \emph{Proc. 46th IEEE Conf.
  on Decis. and Cont.}, 2007, pp. 2289--2294.

\bibitem{leng2011distributed}
M.~Leng and Y.-C. Wu, ``Distributed clock synchronization for wireless sensor
  networks using belief propagation,'' \emph{{IEEE} Trans. Signal Process.},
  vol.~59, no.~11, pp. 5404--5414, 2011.

\bibitem{etzlinger2013cooperative}
B.~Etzlinger, H.~Wymeersch, and A.~Springer, ``Cooperative synchronization in
  wireless networks,'' \emph{{IEEE} Trans. Signal Process.}, vol.~62, no.~11,
  pp. 2837--2849, 2014.

\bibitem{denis2006joint}
B.~Denis, J.-B. Pierrot, and C.~Abou-Rjeily, ``Joint distributed
  synchronization and positioning in uwb ad hoc networks using toa,''
  \emph{{IEEE} Trans. Microw. Theory Tech.}, vol.~54, no.~4, pp. 1896--1911,
  2006.

\bibitem{zheng2010joint}
J.~Zheng and Y.-C. Wu, ``Joint time synchronization and localization of an
  unknown node in wireless sensor networks,'' \emph{{IEEE} Trans. Signal
  Process.}, vol.~58, no.~3, pp. 1309--1320, 2010.

\bibitem{zhu2010joint}
S.~Zhu and Z.~Ding, ``Joint synchronization and localization using toas: a
  linearization based wls solution,'' \emph{{IEEE} J. Sel. Areas Commun.},
  vol.~28, no.~7, pp. 1017--1025, 2010.

\bibitem{huang2013efficient}
J.~Huang, Y.~Xue, and L.~Yang, ``An efficient closed-form solution for joint
  synchronization and localization using toa,'' \emph{Future Gener. Comp. Sy.},
  vol.~29, no.~3, pp. 776--781, 2013.

\bibitem{bancroft1985algebraic}
S.~Bancroft, ``An algebraic solution of the gps equations,'' \emph{{IEEE}
  Trans. Aerosp. Electron. Syst.}, no.~1, pp. 56--59, 1985.

\bibitem{ahmad2013joint}
A.~Ahmad, E.~Serpedin, H.~Nounou, and M.~Nounou, ``Joint node localization and
  time-varying clock synchronization in wireless sensor networks,'' in
  \emph{Proc. IEEE Int. Conf. on Acous., Speech and Sig. Process.}, 2013, pp.
  5170--5174.

\bibitem{meyer2013distributed}
F.~Meyer, B.~Etzlinger, F.~Hlawatsch, and A.~Springer, ``A distributed
  particle-based belief propagation algorithm for cooperative simultaneous
  localization and synchronization,'' in \emph{Proc. Asilomar Conf. Sig.,
  Syst., Comput.}, 2013, pp. 527--531.

\bibitem{etzlinger2013cooperative1}
B.~Etzlinger, F.~Meyer, A.~Springer, F.~Hlawatsch, and H.~Wymeersch,
  ``Cooperative simultaneous localization and synchronization: A distributed
  hybrid message passing algorithm,'' in \emph{Proc. Asilomar Conf. Sig.,
  Syst., Comput.}, 2013, pp. 1978--1982.

\bibitem{gezici2004uwb}
S.~Gezici and Z.~Sahinoglu, ``Uwb geolocation techniques for ieee 802.15. 4a
  personal area networks,'' \emph{Mitsubishi Electric Research Laboratory
  Technical Report TR-2004-110}, 2004.

\bibitem{van2012comparison}
S.~Van~de Velde, H.~Wymeersch, and H.~Steendam, ``Comparison of message passing
  algorithms for cooperative localization under nlos conditions,'' in
  \emph{Proc. 9th Workshop Position. Navi. Commun.}\hskip 1em plus 0.5em minus
  0.4em\relax IEEE, 2012, pp. 1--6.

\bibitem{wymeersch2012machine}
H.~Wymeersch, S.~Maran{\`o}, W.~M. Gifford, and M.~Z. Win, ``A machine learning
  approach to ranging error mitigation for uwb localization,'' \emph{{IEEE}
  Trans. Commun.}, vol.~60, no.~6, pp. 1719--1728, 2012.

\bibitem{liu2013analysis}
D.~Liu, M.-C. Lee, C.-M. Pun, and H.~Liu, ``Analysis of wireless localization
  in nonline-of-sight conditions,'' \emph{{IEEE} Trans. Veh. Technol.},
  vol.~62, no.~4, pp. 1484--1492, 2013.

\bibitem{pearl1986fusion}
J.~Pearl, ``Fusion, propagation, and structuring in belief networks,''
  \emph{Artif. intell.}, vol.~29, no.~3, pp. 241--288, 1986.

\bibitem{winn2005variational}
J.~Winn, C.~M. Bishop, and T.~Jaakkola, ``Variational message passing.''
  \emph{J. Mach. Learn. Res}, vol.~6, no.~4, 2005.

\bibitem{sathyan2011wasp}
T.~Sathyan, D.~Humphrey, and M.~Hedley, ``Wasp: A system and algorithms for
  accurate radio localization using low-cost hardware,'' \emph{{IEEE} Trans.
  Syst., Man, Cybern. {C}}, vol.~41, no.~2, pp. 211--222, 2011.

\bibitem{loeliger2004introduction}
H.-A. Loeliger, ``An introduction to factor graphs,'' \emph{{IEEE} Signal
  Process. Mag.}, vol.~21, no.~1, pp. 28--41, 2004.

\bibitem{buntine1994operations}
W.~L. Buntine, ``Operations for learning with graphical models,'' \emph{J.
  Artif. Int. R}, vol.~2, pp. 159--225, 1994.

\bibitem{chen2006network}
J.-C. Chen, Y.-C. Yeong-Cheng~Wang, C.-S. Maa, and J.-T. Chen, ``Network-side
  mobile position location using factor graphs,'' \emph{{IEEE} Trans. Wireless
  Commun.}, vol.~5, no.~10, pp. 2696--2704, Oct 2006.

\bibitem{aldershof1995facts}
B.~Aldershof, J.~Marron, B.~Park, and M.~Wand, ``Facts about the gaussian
  probability density function,'' \emph{Appl. Anal.}, vol.~59, no. 1-4, pp.
  289--306, 1995.

\bibitem{rockafellar1998variational}
R.~T. Rockafellar, R.~J.-B. Wets, and M.~Wets, \emph{Variational
  analysis}.\hskip 1em plus 0.5em minus 0.4em\relax Springer, 1998, vol. 317.

\bibitem{dauwels2007variational}
J.~Dauwels, ``On variational message passing on factor graphs,'' in \emph{Proc.
  IEEE Int. Symp. on Inf. Theory}, 2007, pp. 2546--2550.

\bibitem{hotta2003mean}
T.~Hotta, ``Mean-field approximation,'' in \emph{Nanoscale Phase Separation and
  Colossal Magnetoresistance}.\hskip 1em plus 0.5em minus 0.4em\relax Springer,
  2003, pp. 157--167.

\bibitem{yedidia2005constructing}
J.~S. Yedidia, W.~T. Freeman, and Y.~Weiss, ``Constructing free-energy
  approximations and generalized belief propagation algorithms,'' \emph{{IEEE}
  Trans. Inf. Theory}, vol.~51, no.~7, pp. 2282--2312, 2005.

\bibitem{riegler2013merging}
E.~Riegler, G.~E. Kirkelund, C.~N. Manch{\'o}n, M.~Badiu, and B.~H. Fleury,
  ``Merging belief propagation and the mean field approximation: A free energy
  approach,'' \emph{{IEEE} Trans. Inf. Theory}, vol.~59, no.~1, pp. 588--602,
  2013.

\bibitem{opper2001advanced}
M.~Opper and D.~Saad, \emph{Advanced mean field methods: Theory and
  practice}.\hskip 1em plus 0.5em minus 0.4em\relax MIT press, 2001.

\bibitem{lien2012comparison}
J.~Lien, U.~J. Ferner, W.~Srichavengsup, H.~Wymeersch, and M.~Z. Win, ``A
  comparison of parametric and sample-based message representation in
  cooperative localization,'' \emph{Int. J. Nav. Obser.}, vol. 2012, 2012.

\end{thebibliography}
\end{document}